\documentclass[aip,graphicx,
reprint,
superscriptaddress,
amsmath,amssymb,
aps,
prl,
]{revtex4-1}

\usepackage[latin1]{inputenc}
\usepackage{amsmath}
\usepackage{amsfonts}
\usepackage{hyperref}
\usepackage{amssymb}
\usepackage{graphicx}
\usepackage{hyperref}
\usepackage{colortbl}
\usepackage{subfigure}
\usepackage{gensymb}
\usepackage{svg}

\usepackage[T1]{fontenc}
\usepackage{mathptmx}
\usepackage{bbold}
\usepackage{scalerel}
\usepackage{xcolor}
\usepackage{mathtools}
\usepackage{epstopdf}

\newcommand{\hyref}[1]{\hyperref[#1]{\ref{#1}}}
\newcommand{\dd}{\mathrm{d}}

\newcommand{\orange}[1]

\begin{document}
\title{Pressure-sensitive ion conduction in a conical channel: optimal pressure and geometry }
	\author{W.Q. Boon}
	\thanks{These two authors contributed equally}
	\affiliation{Institute for Theoretical Physics, Utrecht University,  Princetonplein 5, 3584 CC Utrecht, The Netherlands}
	\author{T.E. Veenstra}
	\thanks{These two authors contributed equally}
		\affiliation{Institute for Theoretical Physics, Utrecht University,  Princetonplein 5, 3584 CC Utrecht, The Netherlands}
	\author{M. Dijkstra}
	\affiliation{Soft Condensed Matter, Debye Institute for Nanomaterials Science, Utrecht University, Princetonplein 1, 3584 CC Utrecht, The Netherlands}
	\author{R. van Roij}
		\affiliation{Institute for Theoretical Physics, Utrecht University,  Princetonplein 5, 3584 CC Utrecht, The Netherlands}

\date{\today}

\begin{abstract}
Using both analytic and numerical analyses of the Poisson-Nernst-Planck equations we theoretically investigate the electric conductivity of a conical channel, which in accordance with recent
experiments exhibits a strong non-linear pressure dependence. This mechanosensitive diodic behavior stems
from the pressure-sensitive build-up or depletion of salt in the pore. From our analytic results we find that the optimal geometry for this diodic behavior strongly depends on the flow rate, the ideal ratio of tip-to-base-radii being equal to 0.22 at zero flow. With increased flow this optimal ratio becomes smaller and simultaneously the diodic performance becomes weaker. Consequently an optimal diode is obtained at zero-flow, which is realized by applying a pressure drop that is proportional to the applied potential and to the inverse square of the tip radius thereby countering electro-osmotic flow. When the applied pressure deviates from this ideal pressure drop the diodic performance falls sharply, explaining the dramatic mechanosensitivity observed in experiments. 
\end{abstract}

\maketitle
A fluidic channel allows for the simultaneous transport of solvent, charge, and dissolved salt when connected to two liquid electrolyte reservoirs at different pressures, voltages, salt concentrations, and/or temperatures. Such ionic transport is not only interesting from a fundamental point of view, but also for energy harvesting \cite{technology1,bluepower1,bluepower2,bluepower3}, desalination \cite{desal1,desal2} and microfluidic applications \cite{microfluidics1,microfluidics2}. In all these devices fluidic channels with dimensions in the nano- and micrometer regime are used \cite{nanofluidics1,nanofluidics2,nanofluidics3}, a size range where the influence of surface charge on transport becomes significant due to the relatively large surface-to-volume ratio. This surface charge is key to electrokinetic transduction phenomena such as the flow of electrolyte by an electric potential drop (electro-osmosis) or the electric (streaming) current induced by flow due to an applied pressure drop \cite{surf1,surf2,surf3}. While these electrokinetic transduction phenomena have long been understood\cite{levine1975theory,linEK1,linEK2,linEK3}, at least in simple channel geometries, in conical pores exotic transport behavior such as electro-osmotic flow inversion \cite{flowreverse1,flowreverse2,flowreverse3}, other non-linear flow-effects \cite{asgharflow1,asgharflow2} and current rectification \cite{rect1,rect2,rect3, woermann1, woermann2,rect1largeR, rect2largeR} have been observed. Such non-linear transport behavior makes conical pores uniquely attractive for biochemical sensing \cite{biosensor1,biosensor2,biosensor3, biosensorsreview,biochemphysoffluids, biochemPOF1, biochemPOF2,biochemPOF3} and neuromorphic applications \cite{biotransport1, conereview, technology3,AIPmem}. 
In this letter, we analyse the intricate case of a micron-sized cone-shaped channel exposed to a simultaneous pressure and electric potential drop by means of the well-known Poisson-Nernst-Planck-Stokes (PNPS) equations. We will show that the ionic current in conical nanopores can be either strongly reduced or enhanced by a pressure difference and concomitant flow, resulting into an extremely mechanosensitive ionic diode similar to those present in cell membranes  \cite{mechanosensing1,mechanosensing2} and such a pressure-sensitivity can also be used to optimize power generation in artificial pores\cite{qian}. Recent experiments revealed such a non-linear pressure-induced electric transport in conical pores even at micrometer length scales \cite{jubin}. It was found that the electric current $I(\Delta P, \Delta \psi)$ due to an applied potential difference $\Delta \psi$ is very sensitively dependent on the applied pressure drop $\Delta P$ over the channel. 
\begin{figure}[b!]
    \includegraphics[width=0.9\linewidth]{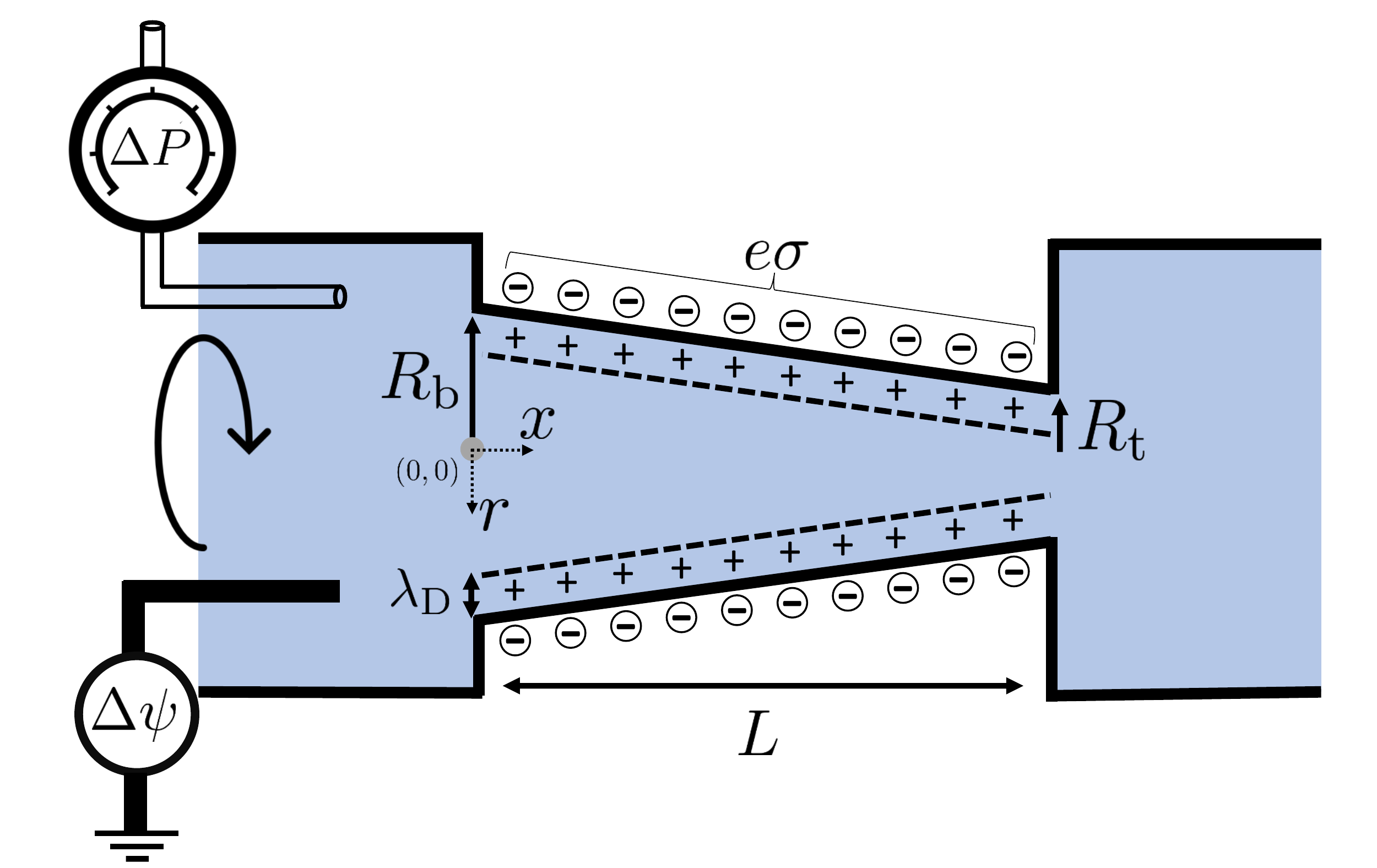}
    \caption{Schematic of an axially symmetric conical channel of length $L$, base radius $R_{\rm{b}}$ at $x=0$, and tip radius $R_{\rm{t}}\leq R_{\rm{b}}$ at $x=L$. The channel connects two bulk 1:1 electrolytes in the half spaces $x<0$ and $x>L$. The channel wall carries a negative surface charge density $e\sigma$ that induces an electric double layer of thickness $\lambda_{\rm{D}}$, the Debye length. Volume, charge, and salt ions are transported through the channel by an applied potential drop $\Delta\psi$ and pressure drop $\Delta P$.  }
     \label{schematic}
\end{figure}
Surprisingly, the observed pressure dependence of the electric conductance occurred at extremely low rather than high pressures. For conical pores it was already observed that, for $\Delta P=0$, the response of the current $I$ is asymmetric with regard to the sign of $\Delta\psi$ and this so-called current rectification is attributed to concentration polarization \cite{rect1,rect2,rect3, rect1largeR, rect2largeR, woermann1, woermann2,curRecPOF}. Here we show that the flow (and hence pressure)-sensitive conductivity for $\Delta P\neq 0$ can also be understood by the concentration-polarization, in contrast to earlier work which suggests that novel mechanisms are needed such as a bulk space-charge\cite{jubin} or a non-linear streaming current\cite{qian}. Such a flow-sensitive conduction was previously noted in numerical calculations, which however ignored electro-osmotic flow\cite{whitepressure} that we find to be of great importance. Whether electro-osmotic flow can\cite{whiteNoEOflow, halfangle, siwyNoEOflow, halfangle3noflow, AIPnoflow} or cannot\cite{hsuEOflow1, hsuEOflow2, aiEOflow, wangEOflow} be ignored is debated in the literature and we reconcile these two opposing views by showing that the importance of flow depends on P\'eclet number; in the small P\'eclet regime\cite{whiteNoEOflow}  flow can be ignored while in the large P\'eclet regime\cite{hsuEOflow1} it is important. This large-Pe regime is natural for large micrometer channels\cite{rect1largeR, rect2largeR, aartscone, jungyul, jubin} common in experiments, while theoretical descriptions of such systems often neglect flow\cite{cengio, tradeoff, no-flow-ICR, halfangle3noflow}. To account for the effect of flow on rectification we derive for the first time a closed-form expression for the ion-distribution in the channel. Together with the $2\times 2$  transport matrix accounting for both the pressure-driven flow and the electro-osmotic flow neglected by Ref.\cite{whitepressure} this concentration profile naturally accounts for the mechanosensitive cone conductance, without needing to invoke any coupling between current and the Maxwell stress-tensor as was done in Ref.\cite{jubin}. Such a pressure-sensitivity cannot be captured by recent analytic theories as they neglect flow entirely\cite{cengio,tradeoff}. Furthermore we find that the optimal cone geometry for rectification strongly depends on the flow rate, while studies searching numerically for such a geometry often neglect exactly this feature \cite{halfangle,no-flow-ICR, halfangle3noflow}. 

We consider two reservoirs of an aqueous 1:1 electrolyte in the two half spaces $x<0$ and $x>L$ connected by an axially-symmetric cone-shaped channel of length $L$. Here $x$ is the cartesian coordinate along the symmetry axis; the radial coordinate is $r$. The channel has a wide base radius $R_{\rm{b}}\ll L$ at $x=0$ and a narrow tip radius $R_{\rm{t}}\leq R_{\rm{b}}$ at $x=L$. The radius of the channel for $x\in[0,L]$ reads $R(x)= R_{\rm{b}}- (x/L)(R_{\rm{b}}-R_{\rm{t}})$. The channel wall at $r=R(x)$ carries a uniform negative surface charge density $e\sigma$, with $e$ the proton charge. The two reservoirs both contain an identical aqueous 1:1 electrolyte with viscosity $\eta$, ionic diffusion coefficient $D$, dielectric permittivity $\epsilon$, and total ionic bulk concentration $2\rho_{\rm{b}}$. Thus, asymptotically far from the channel, at either side $|x|/L\gg1$, the local cation concentration $\rho_+(x,r)$ and anion concentration $\rho_-(x,r)$ are both equal to $\rho_{\rm{b}}$. 

Inspired by the experiments of Ref. \cite{jubin} we consider an applied pressure drop $\Delta P$ and a simultaneous electric potential drop $\Delta\psi$ across the channel. These steady driving forces give rise to a potential $\psi(x,r)$ and a pressure excess $P(x,r)-P_0$ which vanish in the bulk phase $x\gg L$ and are equal to $\Delta\psi$ and $\Delta P$, respectively, for $x\ll -L$, where $P_0$ is an arbitrary reference pressure. They drive a fluid flow with velocity $\mathbf{u}(x,r)$ and ionic fluxes ${\bf j}_{\pm}(x,r)$, leading to nontrivial concentration profiles $\rho_{\pm}(x,r)$. 
In the Supplementary Material I (SM I) we present the standard Poisson-Nernst-Planck-Stokes (PNPS) equations and the blocking and no-slip boundary conditions. Together with Gauss' law for the surface charge, they form a closed set for ${\bf u}$, $\psi$, ${\bf j}_{\pm}$, and $\rho_\pm$.  Convenient linear combinations are the total salt concentration $\rho_\mathrm{s}=\rho_\mathrm{+}+\rho_\mathrm{-}$, the charge density $\rho_\mathrm{e}=\rho_\mathrm{+}-\rho_\mathrm{-}$, and the associated fluxes $\mathbf{j}_{\mathrm{s}}=\mathbf{j}_++\mathbf{j}_-$ and 
$\mathbf{j}_{\mathrm{e}}=\mathbf{j}_+-\mathbf{j}_-$. 
In equilibrium, i.e. for vanishing $\Delta P$ and $\Delta\psi$, all fluxes vanish and the PNPS equations describe an Electric Double Layer (EDL) with an excess of cations and a depletion of anions close to $r=R(x)$ such that the negative surface charge is compensated  \cite{physicalchem}. The thickness of the EDL is given by the Debye length $\lambda_{\rm{D}}=\sqrt{\epsilon k_\mathrm{B}T/2e^2\rho_{\rm{b}}}=10$ nm for the case $\rho_{\rm{b}}=1$mM that we consider. 

Inspired by the experimental conditions of Ref. \cite{jubin}, the focus of this letter will be on the long-channel thin-EDL limit with $L\gg R_{\rm{b}}\geq R_{\rm{t}}\gg\lambda_{\rm{D}}$ such that EDL-overlap does not play a role. This is in contrast to a large body of literature on non-linear transport in cone-shaped channels, where overlap of the EDL is a key ingredient for current rectification and diodic behavior \cite{rect1, rect2,rect3}.  We will show that the conical geometry combined with simultaneous pressure- and potential-induced transport leads to an $x$-independent volumetric flow rate $Q=2\pi\hat{{\bf x}}\cdot\int_0^{R(x)}{\bf u}(x,r)r dr$ and electric current $I=2\pi e\hat{{\bf x}}\cdot\int_0^{R(x)}{\bf j}_{\rm{e}}(x,r)r dr$
that satisfy an Onsager-like relation
\begin{equation}
\begin{pmatrix}
    Q\\
    I
    \end{pmatrix}=
      \frac{\pi R_{\rm{b}}R_{\rm{t}}}{L}
    \begin{pmatrix}
     \mathbb{L}_{11} & \mathbb{L}_{12} \\[6pt] 
     \mathbb{L}_{21} & \mathbb{L}_{22}(\Delta P,\Delta \psi)
    \end{pmatrix}
    \begin{pmatrix}
    \Delta P\\
    \Delta \psi
    \end{pmatrix}.
    \label{Onsager1}
\end{equation}
We set out to calculate all elements of the transport matrix $\mathbb{L}$ analytically, not only the permeability $\mathbb{L}_{11}$ and the electro-osmotic mobility $\mathbb{L}_{12}=\mathbb{L}_{21}$, but also the  electric conductance $\mathbb{L}_{22}$ that, as we will see, strongly depends on the applied pressure-and voltage drop -in agreement with experiments \cite{jubin}. This pressure sensitivity is due to highly nontrivial ion concentration profiles that vary on length scales of the channel dimensions, as follows from our analytic expression obtained from the PNPS equations. From this we will find that optimal current rectification requires not only a pressure drop $\Delta P^*=-\mathbb{L}_{12}\Delta\psi/\mathbb{L}_{11}$ (such that $Q=0$) but also a universal optimal geometry with $R_{\rm{t}}/R_{\rm{b}}\simeq 0.22$. 
\begin{figure}[t!]
   \includegraphics[width=0.9\linewidth]{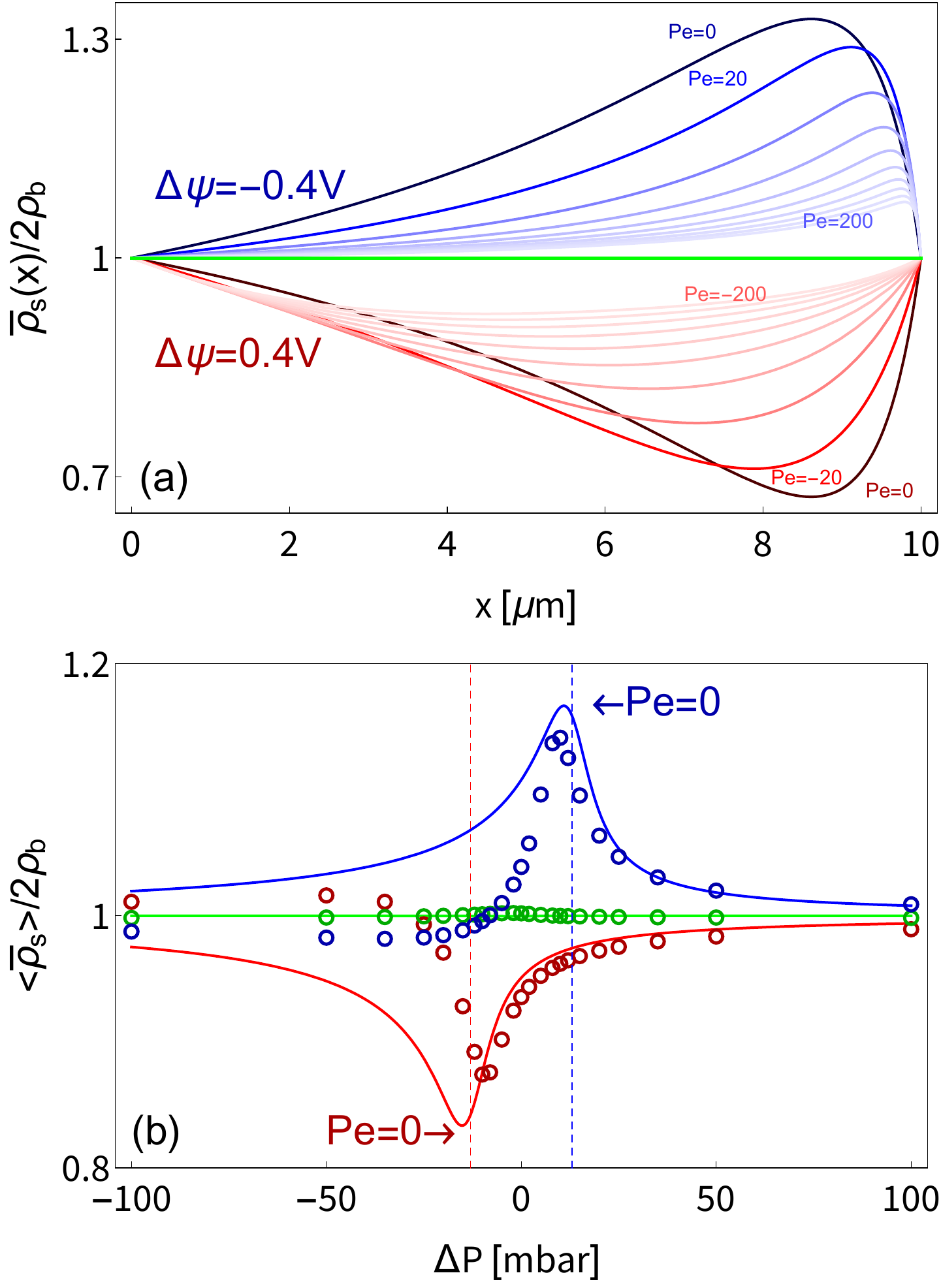}
  \caption{(a) Cross-sectional averaged salt concentration $\bar{\rho}_{\rm{s}}(x)$ normalized by the bulk concentration $2\rho_{\rm{b}}$ as a function of the lateral position $x$ for our standard parameter set (see text). For potential drops $\Delta\psi=+0.4$V (red) and $-0.4$V (blue), for which $\Delta\rho=\mp31$mM according to Eq.~(\ref{drho}), the solid lines represent  concentration profiles at P\'eclet numbers that vary between 0 and $\mp200$ in steps of 20. The green curve represents the case $\Delta\psi=0$V at any Pe. (b) The normalized laterally averaged concentration $\langle\bar{\rho}_{\rm{s}}\rangle/2\rho_{\rm{b}}$ as a function of the pressure drop $\Delta P$ at potential drops $\Delta\psi=+$0.4V (red), $-0.4$V (blue), and $0$V (green). Solid lines represent Eq.~(\hyref{eq:concentration}), data points are from numerical solutions to the full PNPS equations, which for $\Delta\psi=\pm0.4$V show an extremum very close to $\Delta P^*=\mp13$mbar from Eq.~(\hyref{deltapstar}) where Pe=0, denoted by the vertical dashed lines.}
\label{fig:concentration2}
\end{figure}
We solve the PNPS equations for a wide variety of system parameters and  show illustrative examples for the standard parameter set inspired by Ref. \cite{jubin} with tip radius $R_{\rm{t}}=0.17\mu$m, channel length $L=10\mu$m, base radius $R_{\rm{b}}=1.04\mu$m, viscosity $\eta=1$ mPas, dielectric constant 80 times vacuum permittivity, ionic diffusion coefficient $D=1$ nm$^2$/ns, and surface charge $e\sigma=-0.02e/$ nm$^2$, which gives at $\rho_{\rm{b}}=1$mM a zeta potential of $\psi_0=-40$mV corresponding to a silica surface in contact with an aqueous 1:1 electrolyte \cite{silica1}. In line with the Stokes equation we find ${\bf u}(x,r)$ to contain essentially two contributions. (i) A pressure drop on its own induces a Poiseuille-like flow that is directed towards the (virtual) vertex of the cone for $\Delta P>0$, or away from it for $\Delta P<0$   \cite{Qpcone1,Qpcone2,Qpcone3}. Its contribution $ Q_{\rm{P}} \equiv (\pi R_{\rm{b}}R_{\rm{t}}/L)\mathbb{L}_{11}\Delta P$ to $Q$ is independent of $x$ and can be obtained analytically  \cite{Qpcone1,Qpcone2,Qpcone3} to yield $\mathbb{L}_{11}=R_{\rm{b}}^2R_{\rm{t}}^2/8\eta\langle R^2\rangle$, where the angular brackets denote a lateral average  $\langle R^2\rangle=\int_0^L R^2(x) \dd x/L=(R_{\rm{b}}^2+R_{\rm{t}}^2+R_{\rm{b}}R_{\rm{t}})/3$. The excellent agreement between the pressure-drop dependence of our linear expression for $Q_{\rm{P}}$ and our numerically obtained value of $Q$ at $\Delta\psi=0$ is shown in Fig.S1(a) of SM II. (ii) For our negative surface charge the potential drop $\Delta\psi$ on its own induces an electro-osmotic plug-like flow towards the tip of the cone for $\Delta\psi>0$, or away from the tip for $\Delta\psi<0$. We are not aware of an explicit expression in the literature for $\mathbb{L}_{12}$ that characterizes the electro-osmotic flow rate $Q_{\rm{\psi}}\equiv(\pi R_{\rm{b}} R_{\rm{t}}/L)\mathbb{L}_{12}\Delta\psi$ in a conical pore. Here we derive an explicit expression for $\mathbb{L}_{12}$, which  first requires an expression for the cross-sectional averaged electric field $-\partial_x\bar{\psi}(x)$, see Eq. (S1) of SM II, where $\bar{\psi}(x)=2\pi\int_0^{R(x)}\psi(x,r)r\dd r/\pi R^2(x)$. This averaged electric field has to be proportional to the inverse of the cross section $\pi R^2(x)$ in order to be divergence free. 
The proportionality constant
follows, in the long-channel limit, from the condition that $\int_0^L \partial_x\bar{\psi}(x)\dd x=-\Delta\psi$. This yields
\begin{equation}
   \partial_x\bar{ \psi}(x)=-\frac{\Delta\psi}{L}\frac{ R_{\rm{b}} R_{\rm{t}}}{R^2(x)},
    \label{averagepsix}
\end{equation}
which compares well to the numerical  results, as illustrated in Fig. S2 in SM II. Using the standard electro-osmotic mobility $\mathbb{L}_{12}=-\epsilon\psi_0/\eta$ for a cylinder \cite{coupled1}, but now with our laterally varying electric field and radius, we obtain $Q_\psi=\pi R^2(x)(-\epsilon\psi_0/\eta)\partial_x\bar{\psi}(x)$  which with Eq. (\ref{averagepsix}) is independent of $x$ and hence represents a valid divergence-free solution for the stationary state. In Fig. S1(b) of SM II we compare this  expression for $Q_\psi$ as a function of $\Delta\psi$ with numerical calculations. The agreement is good, although minor deviations on the order of $\sim 10\%$ are visible which we attribute to the approximate nature of our $\mathbb{L}_{12}$.

With $\mathbb{L}_{11}$ and $\mathbb{L}_{12}$ established, we continue with $\mathbb{L}_{22}$, for which the total ion concentration $\rho_{\rm{s}}(x,r)$ is expected to play a major role. In our numerical calculations we find weak radial variation of $\rho_{\rm{s}}(x,r)$ outside the EDL-vicinity $r\simeq R(x)$, in agreement with Ref. \cite{whitepressure}.  
Hence within the thin-EDL limit this implies that the cross-sectional averaged concentration $\bar{\rho}_{\rm{s}}(x)$ is a good proxy for the salt concentration at axial position $x$. If we now define the
total salt flux as $J(x)=2\pi\hat{{\bf x}}\cdot\int_0^{R(x)}{\bf j}_{\rm{s}}(x,r)r\dd r$ we can insert the diffusive, conductive, and advective contributions of ${\bf j}_{\rm{s}}$ as given by the PNPS equations in SM I to rewrite the stationarity condition $\partial_x J(x)=0$ for $x\in[0,L]$ as
\begin{equation}\label{J}
    \!\!\!D\partial_x \left(\!\!\pi R^2(x) \partial_x\bar{\rho}_{\rm{s}}(x)\!- \!2\pi R(x) \sigma \frac{e\partial_x\bar{\psi}}{k_\mathrm{B}T}\!\!\right) \!-\! Q\partial_x\bar{\rho}_{\rm{s}}(x)=0.
\end{equation}
Here we use the radial independence of $\rho_{\rm{s}}(x,r)$ and $\psi(x,r)$ in the thin-EDL limit as well as the slab-neutrality condition $2\pi\int_0^{R(x)}\rho_{\rm{e}}(x,r)r\dd r=-2\pi R(x)\sigma$ as derived in SM II. The slab neutrality condition is an important difference with the analysis presented in Ref.\cite{jubin}, where it was suggested that a bulk space charge is of key importance for understanding the observed mechano-sensitivity of conical pores. For a given $\Delta P$ and $\Delta \psi$ we consider $Q$ and $\partial_x\bar{\psi}(x)$ known from Eqs.  (\ref{Onsager1}) and (\ref{averagepsix}), respectively, such that Eq. (\ref{J}) is an ordinary second-order differential equation for $\bar{\rho}_{\rm{s}}(x)$;  together with its solutions presented below it constitutes the key result of this letter. An important role will be played by the conductive contribution $J_{\rm{cond}}(x)$ to $J$ given by $J_{\rm{cond}}(x)= -2\pi D \sigma (e\Delta\psi/k_{\rm{B}}T) R_{\rm{b}}R_{\rm{t}}/R(x)L$, which varies with $x$ in a conical channel and thus acts as a source or sink term in Eq.~(\hyref{J}) that sucks ions into the channel for $\Delta\psi<0$ and pushes them out for $\Delta\psi>0$.

Given the long-channel limit of interest and the equal salinity of both reservoirs, we solve Eq. (\ref{J}) with boundary conditions $\bar{\rho}_{\rm{s}}(0)=\bar{\rho}_{\rm{s}}(L)=2\rho_{\rm{b}}$, resulting in
\begin{subequations}
\begin{gather*}
    \bar{\rho}_{\rm{s}}(x) - 2\rho_{\rm{b}}=  
     \frac{\Delta \rho}{\text{Pe}} \Bigg[
     \frac{x}{L} \frac{ R_{\rm{t}}}{R(x)}-\frac{ \exp\left({\displaystyle\frac{x}{L}\frac{ R_{\rm{t}}^2}{  R_{\rm{b}} R(x)}\text{Pe}}\right)-1}{\exp\left({\displaystyle\frac{R_{\rm{t}}}{ R_{\rm{b}}}\text{Pe}}\right)-1}\Bigg]\tag{4}\\
     =\begin{dcases}
     \frac{\Delta\rho}{2}\frac{x}{L}\big(1-\frac{x}{L}\big) \frac{ R_{\rm{t}}^2}{R^2(x)}  &  \text{if \ \ |Pe|} \ \!  \ll \bigg(\frac{R_{\rm{b}}}{R_{\rm{t}}}\bigg)^2; \\
      \dfrac{\Delta\rho}{2|\text{Pe}|}\bigg(\frac{ R_{\rm{b}}}{R(x)}\big(1-\frac{x}{L}(1+\frac{R_{\rm{t}}}{ R_{\rm{b}}})\big)\mp 1\bigg)& \text{if \ $\pm$Pe}  \gg \bigg(\frac{R_{\rm{b}}}{R_{\rm{t}}}\bigg)^2. 
     \end{dcases}
    \label{eq:concentration}
    \end{gather*}
\end{subequations}
Here we not only introduced the tip P\'eclet number $\text{Pe} \equiv QL/ D\pi  R_{\rm{t}}^2$ with $Q(\Delta P,\Delta \psi)$ given by Eq.~(\hyref{Onsager1}) but also a measure for the concentration inhomogeneity 
\begin{equation}
\Delta \rho \equiv \frac{2(R_{\rm{b}}-R_{\rm{t}})\sigma }{ R_{\rm{t}}^2}\frac{e \Delta \psi}{k_{\rm{B}}T},\label{drho}
\end{equation}
thus $\Delta\rho=0$ if $R_{\rm{b}}=R_{\rm{t}}$ and hence $\bar{\rho}_{\rm{s}}(x)=2\rho_{\rm{b}}$ in this case. 
Note that both $\text{Pe}$ and $\Delta\rho$ have a sign, and that the dependence on the potential drop is not only accounted for by $\Delta\rho$ but also by $\text{Pe}$ through the electro-osmotic contribution to $Q$, see Eq.~(\hyref{Onsager1}). Clearly, Eq.~(\hyref{eq:concentration}) reveals concentration variations on length scales on the order of the full channel length $0\leq x\leq L$, most prominently for smaller $|\text{Pe}|$. Since the P\'eclet number quantifies the importance of flow, we can now reconcile the discrepancy between works which find electro-osmotic flow to be negligible\cite{whitepressure} and others which find it to be important\cite{hsuEOflow1}, as the former concerns a parameter set with small Pe$\simeq 10^{-2} (R_{\rm{b}}/R_{\rm{t}})^2$ and the latter with large Pe$\simeq 3  (R_{\rm{b}}/R_{\rm{t}})^2$. For $\Delta\psi=\pm0.4$V, which for our standard parameter set gives $\Delta\rho=\mp31$mM from Eq.~(\hyref{drho}), we plot the concentration profile $\bar{\rho}_{\rm{s}}(x)$ of 
Eq.~(\hyref{eq:concentration}) in Fig.~\hyref{fig:concentration2}(a) for P\'eclet numbers between 0 and $\mp200$. In Fig.~\hyref{fig:concentration2}(b) we plot the salt concentration $\langle\bar{\rho}_{\rm{s}}\rangle$ laterally averaged over the interval $x\in[0,L]$, which will play a key role in the electric conductivity $\mathbb{L}_{22}$, as a function of the imposed pressure drop $\Delta P$ for the three voltage drops $\Delta\psi=+0.4$V (red), 0V (green), and $-0.4$V (blue), as obtained numerically from solutions of the PNPS equations (symbols) and on the basis of a straightforward numerical integration of 
Eq.~(\hyref{eq:concentration}) (lines). The agreement, although not perfect, is very good especially for $\Delta P>0$. Our Eq.~(\ref{eq:concentration}) not only correctly predicts the increase/decrease compared to $2\rho_{\rm{b}}$ for a negative/positive potential drop but also the non-monotonic dependence on $\Delta P$; the absolute difference with $2\rho_{\rm{b}}$ is largest (and on the order of 30\%) for $\Delta P\simeq\mp10$mbar, which corresponds in both cases to $\text{Pe}\simeq0$. The two vertical dashed lines represent the pressure drop $\Delta P^*=-(\mathbb{L}_{12}/\mathbb{L}_{11})\Delta\psi$, where $Q=0$ and hence $\text{Pe}=0$ on the basis of Eq.~(\ref{Onsager1}), such that the optimal concentration polarisation is to be expected. Collecting our earlier results we find the optimal pressure drop per voltage drop
\begin{equation}\label{deltapstar}
  \frac{\Delta P^*}{\Delta\psi}=\epsilon \psi_0\frac{8(R_{\rm{b}}^2+R_{\rm{b}}R_{\rm{t}}+R_{\rm{t}}^2)}{3R_{\rm{b}}^2R_{\rm{t}}^2}\stackrel{R_{\rm{b}}\gg R_{\rm{t}}}{\simeq}\frac{8\epsilon\psi_0}{3R_{\rm{t}}^2},
\end{equation}
which yields about $-32\text{mbar}/\text{V}$ for our standard parameter set, and about -27 mbar/V for the extremely large tip-base ratios $R_{\rm{b}}\gg R_{\rm{t}}$ generated by the extrusion of a pipette in the experiments of Ref.~\cite{jubin} (if we assume $\psi_0=-40$mV common for silica\cite{silica1}). Clearly, the inverse square scaling of $\Delta P^*$ with $R_{\rm{t}}$ is key to explaining the dramatic pressure sensitivity observed in the experiments \cite{jubin}. In fact, our results suggest even more pressure sensitivity for larger conical channels, e.g. for $R_{\rm t}=10\mu$m we have $\Delta P^*$ in the microbar regime, which can already be exerted by the sound of passing traffic \cite{noise90db,noisepressure}. For cases where $\Delta P\gg \Delta P^*$, concentration polarization is washed out by the flow; variation of current with both pressure and voltage then closely follows Ohmic conduction. As flow suppresses diodic performance, at a static pressure drop $\Delta P\neq \Delta P^*$ rectification can also be increased by lowering the dielectric constant $\epsilon$ or increasing the viscosity $\eta$ thereby lowering the electro-osmotic flow rate while keeping $\Delta\rho$ unchanged.\\
\begin{figure}[t!]
   \includegraphics[width=0.9\linewidth]{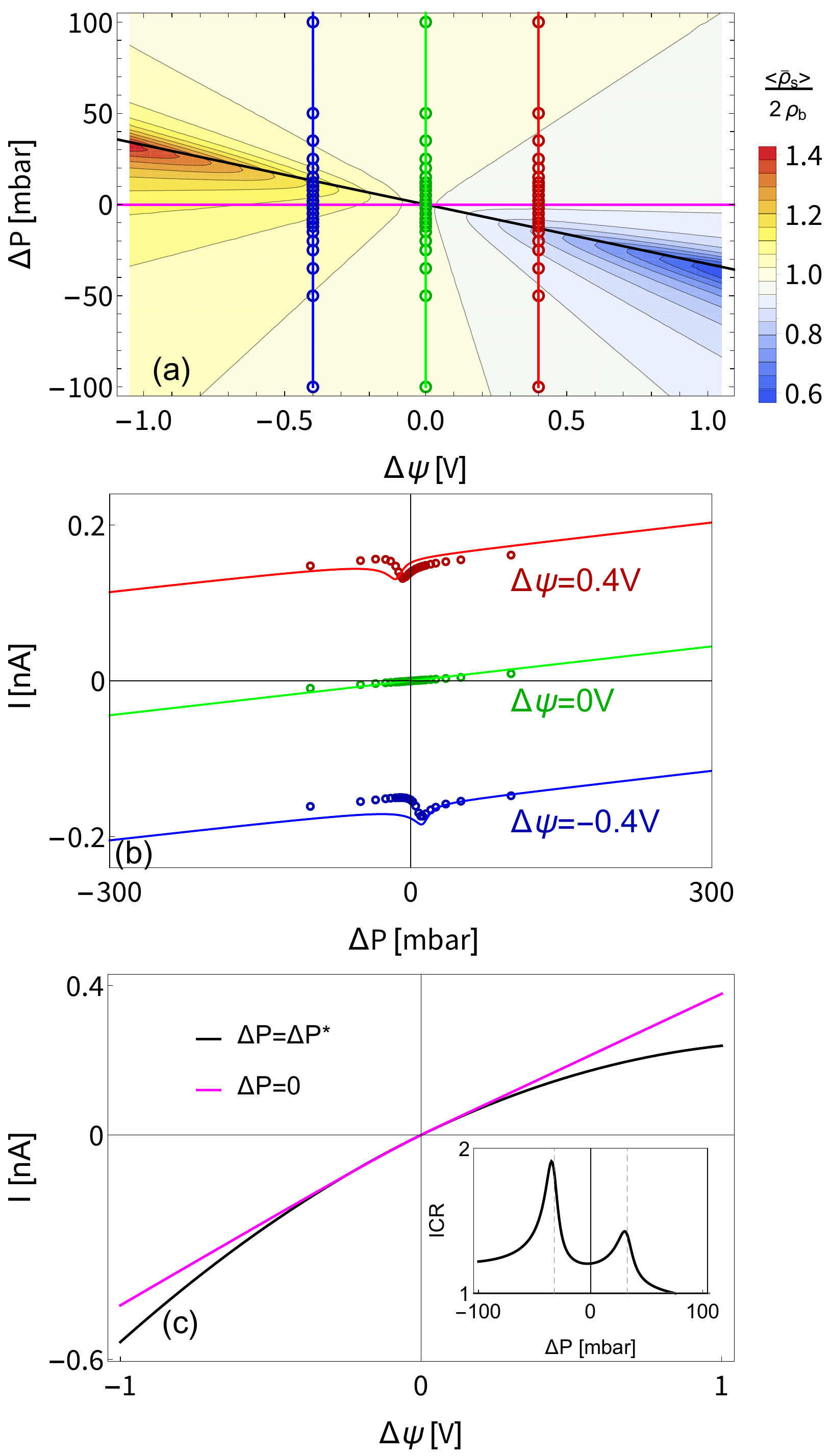}
  \caption{(a) Heat map of the laterally averaged salt concentration $\langle \bar{\rho_{\rm{s}}}\rangle$ for the standard parameter set (see text) in the potential drop $\Delta\psi$ - pressure drop $\Delta P$ plane. (b) Current-pressure ($I-\Delta P$) relation for three fixed potential drops showing a minimum close to $\Delta P=\Delta P^*$ of Eq.~(\hyref{deltapstar}). Symbols represent numerical solutions to the full PNPS equations
  at parameter combinations shown in (a), solid lines represent our analytic solution based on Eq.~(\hyref{Onsager1}). (c) Current-voltage ($I-\Delta\psi$) relation for $\Delta P=0$ (pink) and the optimal pressure drop $\Delta P=\Delta P^*$ (black) which shows increased current rectification ICR compared to $\Delta P=0$. The inset shows the pressure drop dependence of the ICR=$-I(-1$V)/$I(1$V), which exhibits two maxima at $\Delta P=\pm \Delta P^*$.}
\label{fig:contour}
\end{figure}
Now that we have established that Eq.~(\ref{eq:concentration}) gives a fair account of the salt concentration profile in the channel, we will use it to  approximate $\mathbb{L}_{22}$. In the thin-EDL limit the total current $I$ is dominated by the conductive component $-(De/k_{\rm{B}}T)\rho_{\rm{s}}(x,r)\partial_x\psi(x,r)$ of $\hat{{\bf x}}\cdot e{\bf j}_{\rm{e}}(x,r)$ and cross-sectional integration of this current with Eq.~(\hyref{averagepsix}) and the same thin-EDL limit as before yields $I_{\rm{cond}}(x)=\pi R_{\rm{b}}R_{\rm{t}} eD \bar{\rho}_{\rm{s}}(x) (e\Delta\psi/k_{\rm{B}}TL)$, which manifestly depends on $x$ on the basis of Eq.~(\hyref{eq:concentration}). In steady state this lateral variation of the conductive current must be compensated by diffusive and advective currents and the resulting laterally-invariant current $I$ can be obtained by treating the concentration profile $\bar{\rho}_{\rm{s}}(x)$ as a collection of resistors in series~\cite{coupled1}, such that $\mathbb{L}_{22}=(De^2/k_{\rm{B}}T)\langle\bar{\rho}_{\rm{s}}\rangle$ which reveals that conductance is proportional to the laterally averaged salt concentration. 

For our standard parameter set we plot $\langle\bar{\rho}_{\rm s}\rangle$ in Fig.~\hyref{fig:contour}(a) as a heat map in the $(\Delta\psi,\Delta P)$ plane, including a few iso-concentration contours. We clearly see the largest concentration variations, and hence the largest variations of $\mathbb{L}_{22}$, along the black line that represents $\Delta P=\Delta P^*$ of Eq.~(\ref{deltapstar}). In Fig.~\ref{fig:contour}(b) we plot  the $\Delta P$-dependence of the electric current $I$ (lines) as predicted from Eq.~(\ref{Onsager1}) for three voltage drops ($\pm 0.4$ V and zero), together with full numerical calculations (symbols) at the state points indicated by the color-matching symbols in (a) and Fig.~\hyref{fig:concentration2}(b). The overall agreement is quantitative at $\Delta\psi=0$, which is fully in the linear-response regime, while the nonlinear gross features at $\Delta\psi=\pm 0.4$V, especially at $\Delta P\simeq\Delta P^*\simeq \mp 13 $mbar, are accounted for with reasonable accuracy, the more so at the positive potential drop. In Fig.~\ref{fig:contour}(c) we plot current-voltage relations at pressure drops $\Delta P=0$ and $\Delta P=\Delta P^*$, using the same color coding as in (a). The degree of non-Ohmic behavior, characterised by the ionic current rectification $\text{ICR}=-I(-1V)/I(1V)$ is clearly larger at the optimal pressure drop $\Delta P^*$, which is indeed borne out by the inset which shows the full $\Delta P$ dependence  of ICR, revealing peaks at $\pm\Delta P^*$. 

Finally, using our explicit knowledge of $\langle\bar{\rho}_{\rm{s}}\rangle$ and the full transport matrix of Eq.~(\hyref{Onsager1}) we can explicitly search for an optimal cone geometry at which the deviation from Ohmic conductance is largest. Naively Eq.(\hyref{drho}) suggests that for a large concentration profile the ideal tip-to-base ratio should be small ($R_{\rm{t}}/R_{\rm{b}}\ll 1$), however Eq.(\hyref{eq:concentration}) shows that in this limit the concentration profile becomes localized near the tip resulting in a small channel-averaged concentration change. The ideal pore geometry balances the magnitude and spread of the concentration profile and in SM III we show that this optimum occurs at a universal tip-to-base ratio $R_{\rm{t}}/R_{\rm{b}}\simeq 0.22$ for $|\mbox{Pe}| \lesssim 1$. The optimum ratio for concentration polarisation decreases as the power law $b|\mbox{Pe}|^{-\nu}$ with $b=2.5$ and $\nu=0.9$ for $\mbox{Pe}\geq 10^2$, and $b=0.9$ and $\nu=0.55$ for $\mbox{Pe}\leq-10^2$; for all flow rates the ideal tip-to-base ratio is less than $0.22$.
Interestingly, the ideal pore geometry is independent of the channel length $L$, which follows from our Eq.(\hyref{eq:concentration}) for $\bar{\rho}_{\rm{s}}(x)$ that only depends on $x/L$, such that $\mathbb{L}_{22}$ (and in fact the whole matrix $\mathbb{L}$) is independent of the channel length. Hence the concentration polarization does not depend on the cone opening angle, which is surprising as most authors identify it as the key geometric parameter controlling pressure-sensitivity\cite{jubin} and current rectification\cite{halfangle,halfangle2}. 

In conclusion, we provide a full microscopic understanding of the ultra-sensitive pressure- and voltage dependence of the electric conductivity of cone-shaped channels. We identify, and quantify, concentration polarisation due to geometric frustration which leads to a source term in Eq.~(\hyref{J}), even in the thin-EDL case considered here. Moreover, we found an optimal channel geometry $R_{\rm{t}}/R_{\rm{b}}\simeq 0.22$ and an optimal operation condition Eq.~(\hyref{deltapstar}) for current rectification. These insights are important for further developments of mechanotronic~\cite{mechanosensing1,mechanosensing2, mechanosensing3} and biochemical~\cite{biosensor1,biosensor2,biosensor3} sensing as well as microfluidic~\cite{microfluidics1,microfluidics2} and neuromorphic applications~\cite{biotransport1, biotransport2}.

\begin{acknowledgments}
This work is part of the D-ITP consortium, a program of the Netherlands Organisation for Scientific Research (NWO) that is funded by  the  Dutch  Ministry  of  Education,  Culture  and  Science (OCW).
\end{acknowledgments}

\end{document}


\title{Supplementary Material for Pressure sensitive ion-conduction in conical channel: optimal pressure and geometry }

\author{W.Q. Boon}
	\thanks{These two authors contributed equally}
	\affiliation{Institute for Theoretical Physics, Utrecht University,  Princetonplein 5, 3584 CC Utrecht, The Netherlands}
	\author{T.E. Veenstra}
	\thanks{These two authors contributed equally}
		\affiliation{Institute for Theoretical Physics, Utrecht University,  Princetonplein 5, 3584 CC Utrecht, The Netherlands}
	\author{M. Dijkstra}
	\affiliation{Soft Condensed Matter, Debye Institute for Nanomaterials Science, Utrecht University, Princetonplein 1, 3584 CC Utrecht, The Netherlands}
	\author{R. van Roij}
		\affiliation{Institute for Theoretical Physics, Utrecht University,  Princetonplein 5, 3584 CC Utrecht, The Netherlands}
\date{\today}
\maketitle
\beginsupplement

\section{Poisson-Nernst-Planck-Stokes equation and boundary conditions}
In the main text we introduced an axially symmetric conical channel of length $L$, base radius $R_{\rm{b}}$ at $x=0$, and tip radius $R_{\rm{t}} \leq R_{\rm{b}}$ at $x=L$, with $x$ the cartesian coordinate that runs along the symmetry axis. The channel connects two bulk reservoirs of an aqueous 1:1 electrolyte both at the same ionic bulk concentration $2\rho_b$. The viscosity is $\eta$, the ionic diffusion coefficient is $D$, the dielectric constant is $\epsilon$, and the Debye length is $\lambda_D=\sqrt{\epsilon k_BT/2\rho_be^2}$ where $T$ denotes room temperature, $e$ the elementary charge, and $k_B$ the Boltzmann constant. The channel has a fixed negative surface charge density $e\sigma$ at a radial distance $r=R(x)$ from the central axis, where $R(x)=R_{\rm{b}}-(x/L)(R_{\rm{b}}-R_{\rm{t}})$ for $x\in[0,L]$. We consider transport of solvent, ionic charge, and salt driven by the simultaneous application of a steady potential drop $\Delta\psi$ and a steady pressure drop $\Delta P$. The transport is characterised by a volumetric flow rate $Q$, an electric current $I$, and a salt current $J$ through the channel. Throughout we focus on the long-channel thin-EDL limit $\lambda_D\ll R_{\rm{t}}<R_{\rm{b}}\ll L$, and on the regime of low Reynolds number. We thus ignore overlap of the electric double layers (EDLs) and entrance effects.

In the steady state and the low Reynolds number of interest here, the force balance is given by the steady Stokes equation
\begin{equation}
    \eta\nabla^2 \mathbf{u}= \nabla P+e\rho_{\rm{e}} \nabla\psi,\tag{S1}
    \label{Stokes}\\
\end{equation}
where ${\bf u}(x,r)$ is the fluid velocity, $P(x,r)$ the pressure, $\psi(x,r)$ the electrostatic potential, and $e\rho_{\rm{e}}(x,r)$ the ionic space charge density with $\rho_{\rm{e}}=\rho_+-\rho_-$ and $\rho_{\pm}(x,r)$ the cationic (+) and anionic (-) concentration profile. The first term on the right hand side of Eq.~(\hyref{Stokes}) will mainly be driven by a pressure drop and causes a Poiseuille-like flow through the channel, and the second term represents an electric body force that is mainly driven by a potential drop and causes an electro-osmotic flow as we will see below. In thermodynamic equilibrium, where ${\bf u}={\bf 0}$, the pressure gradient balances the electric body force. 

The ionic fluxes ${\bf j}_{\pm}$ contain a Fickian diffusive, an Ohmic conductive, and Stokesian advective contribution described by the Nernst-Planck equations
\begin{equation}
    {\bf j}_{\pm}=-D\left(\nabla\rho_{\pm} \pm \rho_{\pm}\frac{e\nabla\psi}{k_BT}\right) + \rho_{\pm} {\bf u},\label{NP1}\tag{S2}
\end{equation}
where $D$ is the diffusion coefficient that is assumed to be equal for both ionic species for convenience. In the electrolyte the electric potential satisfies the Poisson equation
\begin{equation}
    \nabla^2\psi=-\frac{e}{\epsilon}\rho_{\rm{e}}, \label{Poisson}
    \tag{S3}
\end{equation}
and on the channel wall, at $r=R(x)$ with $x\in[0,L]$ and surface normal ${\mathbf n}$ pointing into the channel, we impose Gauss law $\mathbf{n}\cdot\nabla\psi=e\sigma/\epsilon$. Note that this form of Gauss  law implicitly assumes that the dielectric constant of the electrolyte is much larger than that of the wall material, such that the electric field lines do not ``leak'' out of the channel. The electrolyte is treated as incompressible, and together with the steady-state of interest this yields the divergence-free flux conditions 
\begin{equation}
    \nabla\cdot \mathbf{u}=0; \quad \nabla\cdot \mathbf{j}_{\pm}=0.\label{divergence-free}
    \tag{S4}
\end{equation}
On the channel walls we also impose no-slip boundary and blocking conditions $\mathbf{u}=0$ and $\mathbf{n}\cdot\mathbf{j}_\pm=0$. 
Deep into the bulk of the reservoir connected to the base, $x\ll-L$, we impose $\rho_\pm=\rho_{\rm{b}}$, $\psi=\Delta\psi$, $P=P_0+\Delta P$ with $P_0$ an arbitrary reference pressure, and deep into the reservoir connected with the tip, $x\gg L$, we impose $\rho_\pm=\rho_{\rm{b}}$, $\psi=0$, and $P=P_0$. 

The analysis of the PNPS equations is greatly facilitated by the linear combinations given by the total local salt concentration $\rho_\mathrm{s}=\rho_\mathrm{+}+\rho_\mathrm{-}$, the ionic charge flux density ${\bf j}_{\rm{e}}={\bf j}_+-{\bf j}_-$, and the salt flux density ${\bf j}_{\rm{s}}={\bf j}_++{\bf j}_-$, in terms of which the Nernst-Planck equations (\ref{NP1}) can be rewritten as
\begin{gather*}
  \mathbf{j}_{\rm{e}}=-D \big(\nabla\rho_{\rm{e}}+\rho_{\rm{s}} \frac{e\nabla\psi}{k_\mathrm{B}T})+\mathbf{u}\rho_{\rm{e}}\label{charge}
  \tag{S5},\\
  \mathbf{j}_{\rm{s}}=-D \big(\nabla\rho_{\rm{s}}+\rho_{\rm{e}} \frac{e\nabla\psi}{k_\mathrm{B}T})+\mathbf{u}\rho_{\rm{s}}\label{salt} . \tag{S6} 
\end{gather*}
Here we note that the conduction terms $\propto\nabla\psi$ are proportional to $\rho_{\rm{s}}$ for the electric flux and to $\rho_{\rm{e}}$ for the salt flux. This coupling will prove to be key to understanding the physics of the cone-shaped channel. In the main manuscript, we refer to Eqs.~(\ref{Stokes})-(\ref{salt}) and their boundary conditions as the Poisson-Nernst-Planck-Stokes (PNPS) equations.

\section{Derivation elements Onsager matrix}
\begin{figure}[b!]
\includegraphics[width=0.9\linewidth]{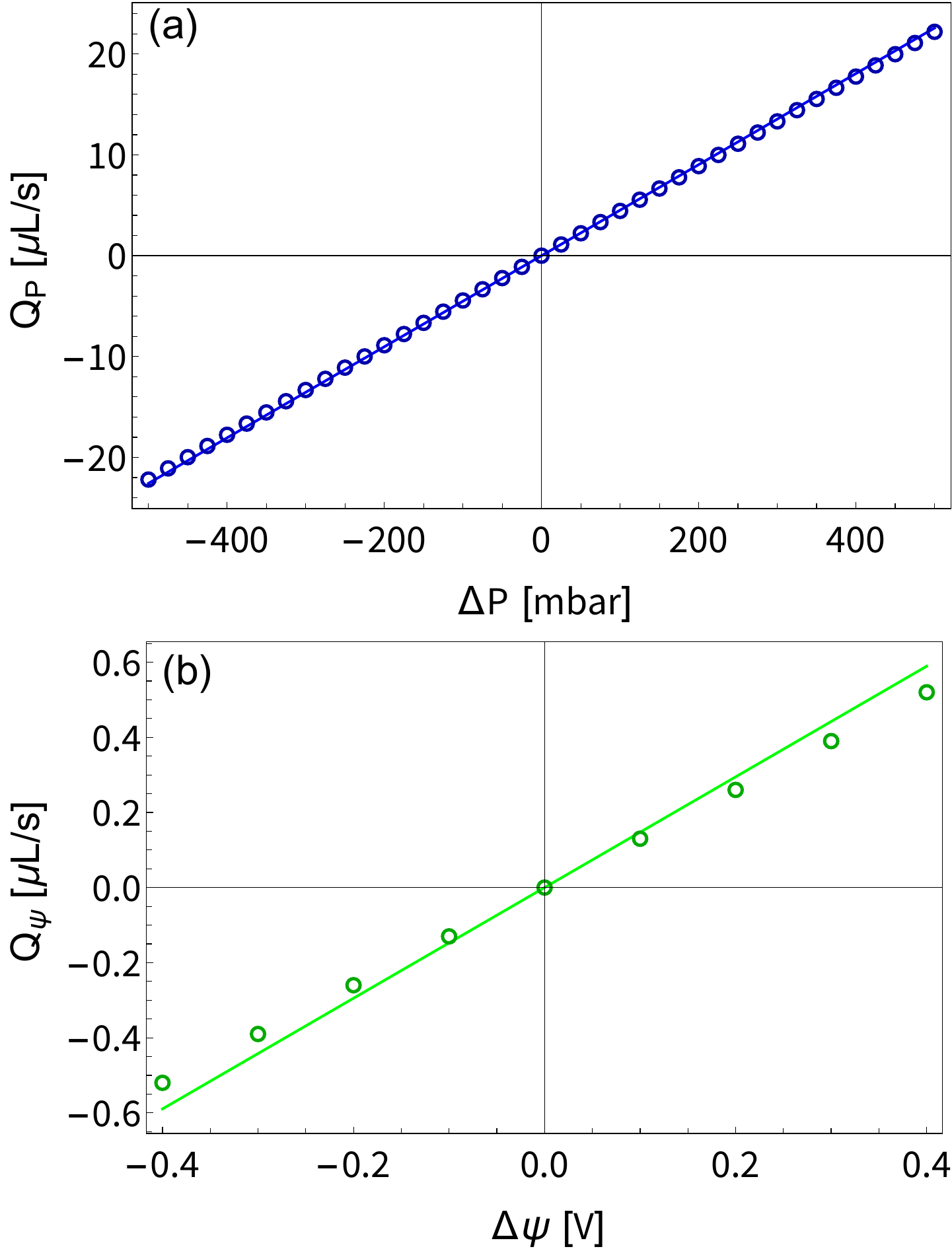}
\caption{(a) Poiseuille-like fluid flux $Q_{\rm{P}}$ as a function of the pressure drop $\Delta P$ at vanishing potential drop $\Delta \psi=0$ and  (b) electro-osmotic potential driven fluid flux $Q_{\rm{\psi}}$ as a function of the potential drop $\Delta\psi$ at a vanishing pressure drop $\Delta P=0$, both for our standard parameter set (see Letter) and obtained from numerical solution of the full PNPS equations (\hyref{Stokes})-(\hyref{salt}) (symbols) and from  $\mathbb{L}_{11}$ and $\mathbb{L}_{12}$ (lines), respectively. Both fluid fluxes are linear in their respective driving force. There is good agreement between analytic and numerical results for the pressure driven flow $Q_{\rm{P}}$, however the analytic expression for the electro-osmotic flow $Q_{\rm{\psi}}$ overestimates the flow rate by $\sim 10\%$.}
\label{fig:fluid}
\end{figure}

\begin{figure}[th!]
\includegraphics[width=0.9\linewidth]{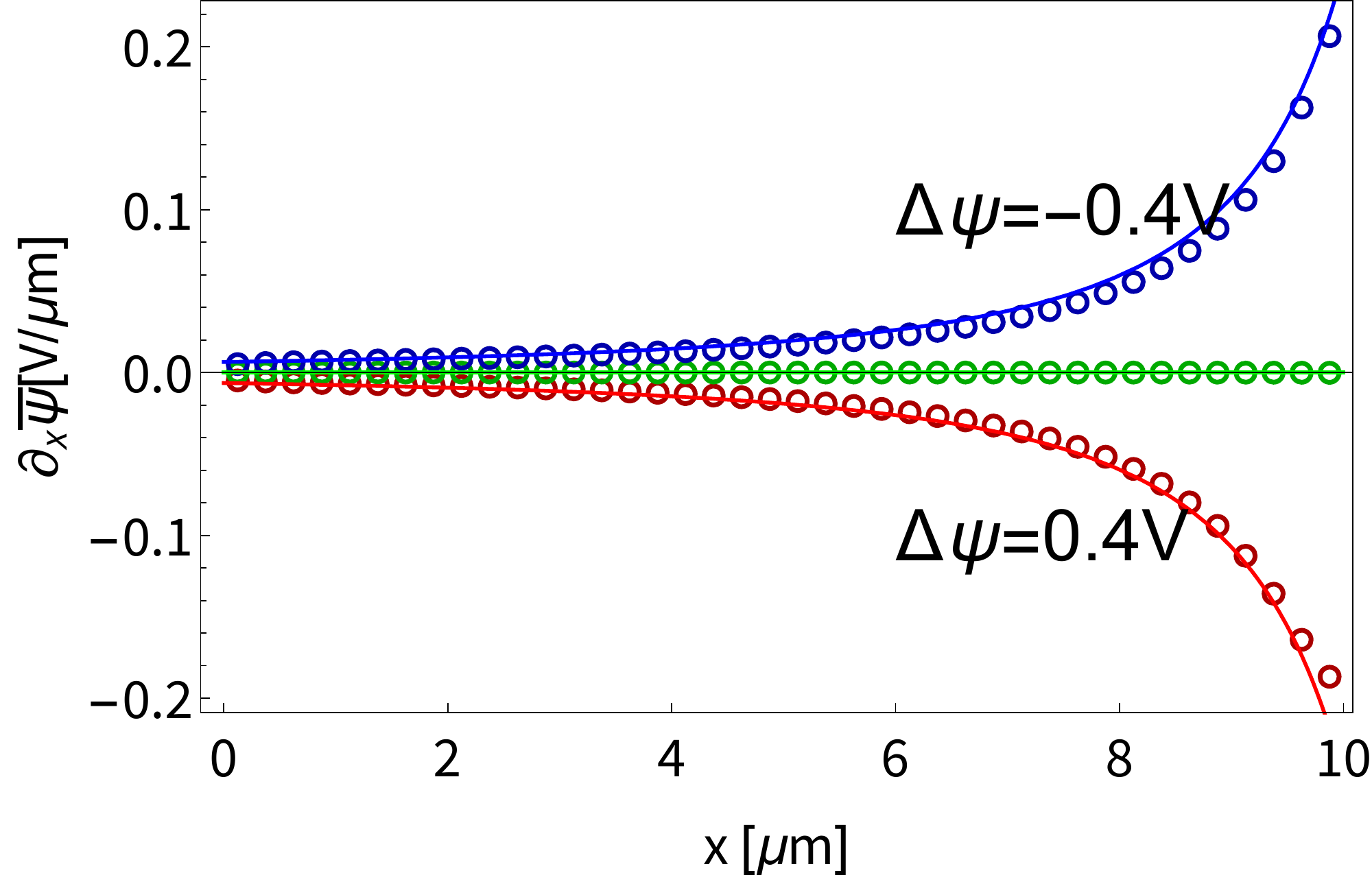}
\caption{The electric field $-\partial_x\bar{\psi}(x)$ for our standard parameter set with pressure drop $\Delta P=0$ as a function of the lateral position $x$ for $\Delta\psi=\pm 0.4$V and $0$ volt (green) obtained from full numeric solutions of the PNPS equations(\hyref{Stokes})-(\hyref{salt}) (symbols) and Eq.~(2) (line) for our standard parameter set (see Letter). Good agreement between numeric and analytic results is found, which confirms the accuracy of Eq.~(2)}.
\label{fig:EvsLength}
\end{figure}

Here we will discuss the details and assumptions involved in the derivation of Eqs.~(1)-(3) of the main text and the components $\mathbb{L}_{11}, \mathbb{L}_{12}$ and $\mathbb{L}_{22}$ of the Onsager-like  matrix starting from the PNPS equations (\hyref{Stokes})-(\hyref{salt}). The hierarchy of length scales $L\gg R_{\rm{b}}\geq R_{\rm{t}} \textgreater \lambda_{\rm{D}}$ serves as the starting point of our derivation. When the channel is much longer than the largest radius $L\gg  R_{\rm{b}}$ entrance-outlet effects to the conductance can be neglected. Additionally, in the long-channel limit all radial components of fluxes and gradients are expected to be much smaller than the corresponding lateral components. Combined with the thin-EDL limit $\lambda_{\rm{D}}\ll R_{\rm{t}}$, which is motivated by the experimental conditions in Ref.~\cite{jubin}, this ensures that the local ion concentrations and the electric field are essentially equal to the cross-sectional averaged salt concentration and electric field, so $\rho_{\rm{s}}(x,r)\simeq\bar{\rho}_{\rm{s}}(x)=2\pi\int_0^{R(x)}\rho_{\rm{s}}(x,r)r\dd r/\pi R^2(x)$  and $\partial_x\psi(x,r)\simeq\partial_x\bar{\psi}(x)=2\pi\int_0^{R(x)}\partial_x\psi(x,r)r\dd r/\pi R^2(x)$.  Moreover, the thin-EDL limit does not only allow us to neglect the influence of channel curvature on the EDL, but also allows us to neglect the influence of salt adsorption~\cite{coupled1} on the cross-sectional averaged salt concentration $\bar{\rho}_{\rm{s}}(x)$. Hence by using the thin-EDL assumption we neglect the inhomogeneous advection of salt through the EDL. For Debye lengths orders of magnitude smaller than the pore radius we expect this assumption to be quite robust, however it will break down at extremely high surface potentials $e\psi_0/k_{\rm{B}}T\gg 1$ as in this regime salt adsorption grows exponentially with $\psi_0$. In summary using the approximations $\rho_{\rm{s}}(x,r)\simeq\bar{\rho}_{\rm{s}}(x)$, $\partial_x\psi(x,r)\simeq \partial_x\bar{\psi}(x)$, $\lambda_{\rm{D}}\ll  R_{\rm{t}}$ in conjunction with the observation from numerical calculations that $|\rho_{\rm{e}}(r\ll R(x))|\ll|\sigma/R(x)|$ will readily result in Eq.~(3) upon radially integrating Eq.~(\hyref{salt}).\\

Before calculating the fluid flux $Q$, we have to verify that the linear response relation Eq.~(1) in the main text is valid for flow, as in the literature there is experimental and numerical evidence that electro-osmotic flow can invert in conical pores under certain experimental conditions~\cite{flowreverse1,flowreverse2,flowreverse3}. This would have a dramatic impact on the pressure sensitivity of the cone. However, as can be seen in Fig.~\hyref{fig:fluid} we find that in the experimental regime of Ref.~[5] no flow inversions occur as $Q$ is linear in both the pressure and potential drop. It should be noted that any non-linearity in the fluid flow $Q(\Delta P,\Delta\psi)$ will significantly change the current-pressure relation $I(\Delta P,\Delta \psi)$. \\
 
 Having verified that the flow $Q$ is linear in their respective driving forces, it now remains to find expressions for first $\mathbb{L}_{11}$ and then $\mathbb{L}_{12}$. When the channel has a tip radius of zero and vanishing surface charge an expression for the fluid flux $Q_{\rm{P}}=(\pi R_{\rm{t}}R_{\rm{b}}/L)\mathbb{L}_{11}\Delta P$ is known~\cite{Qpcone1,Qpcone2,Qpcone3}. Modifying this solution by replacing the pore length $L'$ of a channel with a tip radius of zero with our actual channel length $L= L'(R_{\rm{b}}-R_{\rm{t}})/R_{\rm{b}}$ (with $L'\geq L$) representative of a channel with the same opening angle $2\alpha=2\tan^{-1}(R_{\rm{b}}/L')=2\tan^{-1}((R_{\rm{b}}-R_{\rm{t}})/L)$  but now a non-zero tip radius, we find 
\begin{equation}
     Q_{\rm{P}}(\Delta P)=\dfrac{\Delta P}{\eta} \frac{3\pi L^3  R_{\rm{b}}^3  R_{\rm{t}}^3\alpha^4}{8 (R_{\rm{b}}-R_{\rm{t}})^4(R_{\rm{b}}^2+ R_{\rm{b}}R_{\rm{t}}+R_{\rm{t}}^2)},
    \label{Qp}
    \tag{S7}
\end{equation}
which for $R_{\rm{b}}-R_{\rm{t}}\ll L$ reduces to $\mathbb{L}_{11}$ in the main text. The agreement between this expression for the pressure-driven fluid flux $Q_{\rm{P}}$ (solid line) and the numerically obtained flow (symbols) is remarkable and can be seen in Fig.~\hyref{fig:fluid}(a). \\

We now continue with the calculation of the electro-osmotic flow $Q_\psi$ which first requires an expression for the electric field $-\partial_x\bar{\psi}(x)$ given in the main text by Eq.~(2). This equation is valid under two conditions, (i)  no electric field lines permeate the channel wall, and (ii) the space charge outside of the EDL is negligible. The first condition ensures that all electric field lines remain in the channel and holds when the dielectric constant of the channel wall is much smaller than that of the solvent. The second condition ensures that the divergence of the electric field is zero $\nabla\cdot\nabla\psi(x,r)=0$, for all $r$ several Debye length away from the channel wall, ensuring that no new field lines appear. For straight channels this is a natural assumption, however in conical channels the lateral variation of the electric current $I(x)$ could allow for the build-up of space charge in principle. In our discussion of the numerical results we verify that the effect of this space charge is small and can largely be ignored in the parameter regime of our prime interest. When both condition (i) and (ii) are met the number of electric field lines remains constant over the channel length and the total lateral electric field through a radial slice multiplied by the area of the slice likewise has to be constant, $\pi R^2(x)\partial_x\bar{\psi}(x) =$ constant. Now the electric field as function of lateral position can be found by observing that over the length of the pore the total potential drop has to equal to $\Delta\psi=-\int_0^L\partial_x\bar{\psi} \dd x$, resulting in Eq.~(2) of the Letter. In Fig.~\hyref{fig:EvsLength} we compare $\partial_x\bar{\psi}$ of Eq.~(2) (solid lines), with the numerically obtained function $\partial_x\psi(x,r=0)$ along the symmetry axis in calculations for $\Delta P=0$ (symbols) and find excellent agreement. \\

In order to calculate $\mathbb{L}_{12}$ from the electric field we use the solution for the potential-driven flow through a cylindrical pore but now with position dependent radius and electric field $-\partial_x\bar{\psi}(x)\pi R^2(x)(\epsilon \psi_0/\eta)$ \cite{coupled1} and observe that it is constant over the length of the pore as $\partial_x\bar{\psi}\propto 1/ (\pi R^2(x))$. Hence this Ansatz yields a bonafide divergence-free electro-osmotic flow given by
\begin{equation}
    Q_{\rm{\psi}}(\Delta\psi)=-\frac{\Delta\psi}{L}\pi R_{\rm{t}} R_{\rm{b}} \frac{\epsilon \psi_0}{\eta},
    \label{Qpsi}
\tag{S8}
\end{equation}
where we neglected terms of order $\lambda_{\rm{D}}/R$ on the basis of the thin-EDL limit.  In Fig.~\hyref{fig:fluid}(b) it can be seen that there is a minor deviation of $\sim 10\%$ between $Q_{\rm{\psi}}$ obtained from numerical calculations and this analytic approximation. Implicitly the PNPS equations (\hyref{Stokes})-(\hyref{salt}) allow for diffusio-osmotic fluid flow, driven by the concentration gradient $\partial_x\bar{\rho}_{\rm{s}}$. As the concentration profile and hence the gradient is a non-linear function of the potential drop, any diffusio-osmotic flow would manifest as a non-linear contribution to $Q_\psi(\Delta \psi)$. As no significant deviation from linearity is observed in our numerical results for $Q_{\rm{\psi}}(\Delta\psi)$ we neglect diffusio-osmotic flow.\\

\begin{figure}[b!]
    \includegraphics[width=0.95\linewidth]{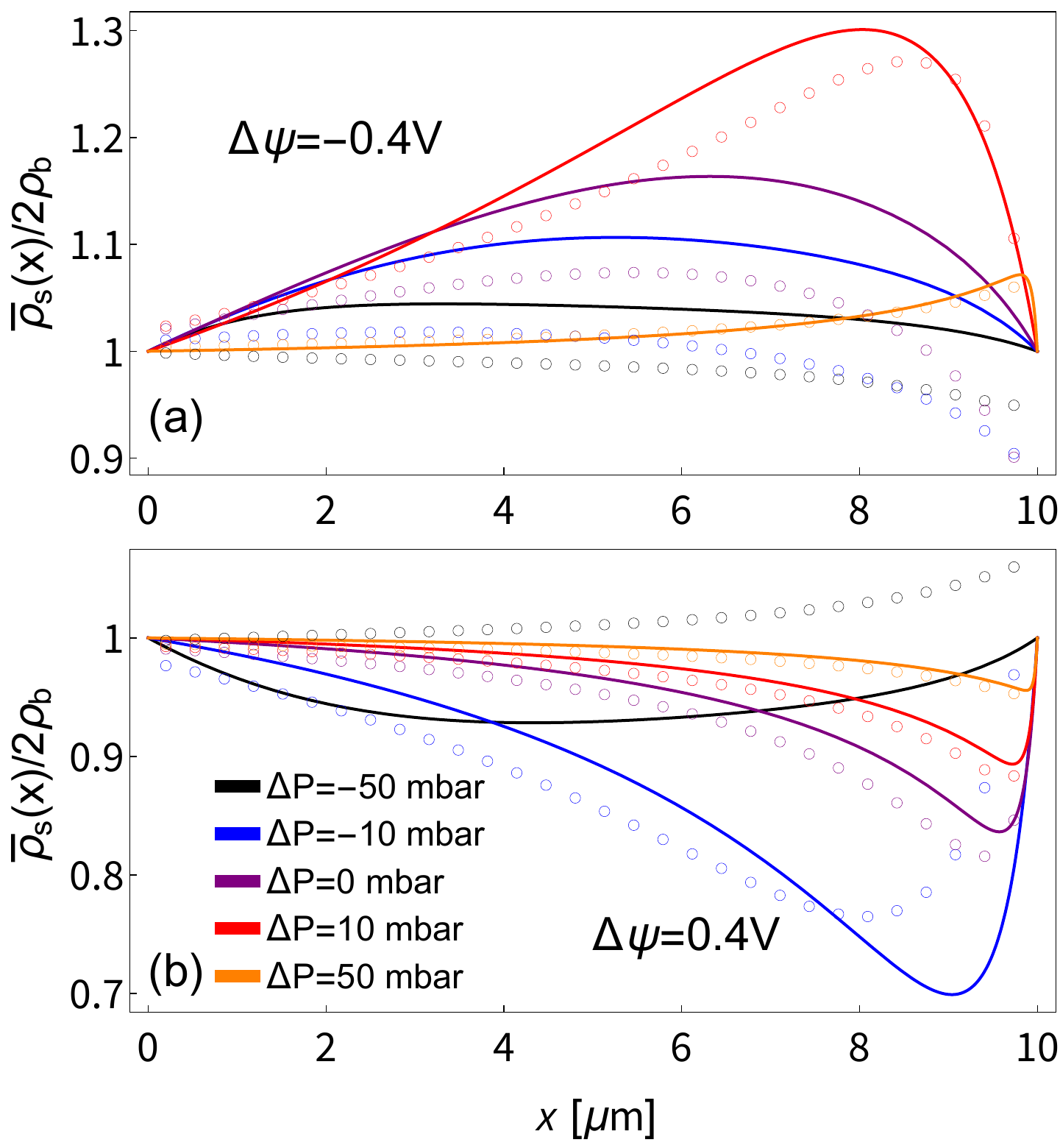}
    \caption{Cross-sectional averaged concentration profile $\bar{\rho}_{\rm{s}}(x)$ over the full channel length from numerical calculations of the full PNPS equations (symbols) compared to curves plotted with Eq.~(4) for our standard parameter set (see main text), potential drop $\Delta\psi=-0.4$V (a) and $\Delta \psi=0.4$V (b) and varying pressure drop $\Delta P$. Numeric and analytic results agree for positive $\Delta P$ but deviate for increasing negative $\Delta P$. Concentration profiles are largest for $\Delta P=\mp 10$ mbar which is rather close to $\Delta P^*=\mp 13$ mbar, where fluid flow vanishes.}
    \label{fig:concentrations}
\end{figure}

Having found expressions for both $\partial_x\bar{\psi}$ and $Q$ we can now straightforwardly solve Eq.~(3), directly yielding Eq.~(4) of the main text. We compare Eq.~(4 (solid lines) with the concentration from numerical calculations in Fig.~\hyref{fig:concentrations} (symbols) for several $\Delta P$ and $\Delta\psi$. We see that while the agreement is not perfect both the non-monotonic $\Delta P$ trend and the overall shape of the concentration profile are captured rather reasonably. It can also be seen that the boundary condition used for analytic calculations ($\rho_{\rm{s}}(0)=\rho_{\rm{s}}(L)=2\rho_{\rm{b}}$) are not fully representative of the numerical calculations, as the concentration profile extends a small distance out of the channel. This discrepancy is possible as in the numerical calculations we apply the boundary condition of bulk concentrations far away rather than at the channel edges. The description of the concentration profiles extending out of the in- and outlets of the conical pore would require a full description of the flow, electric field and currents at the edges of the cone which is not tractable analytically. More significant than the deviation at the tip and base is the sign change of the bulk-excess concentration profile when going from positive $\Delta P$ to $\Delta P\ll -50$mbar, a feature our analytic theory cannot explain. The deviation occurs at very large negative P\'eclet number and represents a secondary non-linearity unrelated to the non-linearity reported in the main text. We speculate the non-linearity may be due to the inhomogeneous advection current $\partial_xI_{\rm{adv}}\propto \partial_x(2Q\sigma/R)\neq 0$ that can build up significant space-charge $\rho_{\rm{e}}(x,r)$ outside the EDL. To study this secondary non-linearity in full detail would require solving for $\partial_xI=0$ and $\partial_xJ=0$ simultaneously. While striking, the opposite sign of our deviation from bulk concentration in our analytic and numeric concentration around Pe$\ll-1$ has little influence on the current $I$, as the concentration deviation is an order of magnitude smaller than the concentration change around Pe = 0. In summary we identify three major sources of error (i) our analytic expression underestimates electro-osmotic flow $Q_{\rm{\psi}}$ by about 10$\%$ (ii) imposing bulk concentrations on the channel edges $\bar{\rho}_{\rm{s}}(0)=\bar{\rho}_{\rm{s}}(L)=2\rho_{\rm{b}}$ implicitly neglects edge effects, and (iii) neglecting minor secondary non-linearities, which are probably related to the lateral variation of the current $I(x)$ that are to be compensated by a (small) space charge distribution $\bar{\rho}_{\rm{e}}(x)$. A final limitation of our theory is that for large negative $\Delta\rho$ the theory allows for $\bar{\rho}_{\rm{s}}(x)$ to become locally negative, which is clearly unphysical. This unphysical result can emerge because the Debye length increases when the concentration decreases, eventually invalidating our starting assumption $\lambda_{\rm{D}}\ll R$. Care should thus be taken not to use the theory in this regime, with a negative conductance a hallmark that the range of validity has been exceeded. Throughout we restrict attention to concentration profiles that deviate less than about 30\% from the bulk concentration, which also allows for equating the relative change of the channel conductance to (the negative of) the change of the relative channel resistance. \\

To calculate the last matrix element $\mathbb{L}_{22}$ we radially integrate Eq.~(\hyref{charge}), resulting in a diffusive, conductive and advective current. In our discussion of numerical results we show that the diffusive current is negligible, which is consistent with the assumption of a negligible space charge outside the EDL, $\rho_{\rm{e}}( r\ll R-\lambda_{\rm{D}},x)\approx 0$. This leaves two components of the current to be calculated $I=I_{\rm{cond}}+I_{\rm{adv}}$ whose ratio scale as $I_{\rm{adv}}/I_{\rm{cond}}\propto \lambda_{\rm{D}}/R$ allowing us to neglect the advective component to $\mathbb{L}_{22}$ when $\lambda_{\rm{D}}/R\ll1$. Now the total current due to a potential drop $\Delta\psi$ is straightforwardly found by integration of the conductive component $-(De/k_BT)\rho_{\rm{s}}(x,r)\partial_x\psi(x,r)$ in Eq.~(\hyref{salt}) and by using $\rho_{\rm{s}}(r,x)\simeq \bar{\rho}_{\rm{s}}(x)$ and $\psi(r,x)\simeq \bar{\psi}(x)$ we find
\begin{equation}
    I_{\rm{cond}}(x)=eD\frac{e\Delta\psi}{k_{\rm{B}}T}\frac{\pi R_{\rm{t}} R_{\rm{b}}}{L} \bar{\rho}_{\rm{s}}(x),
    \label{cond}
    \tag{S9}
\end{equation}
which is inhomogeneous for any non-constant $\bar{\rho}_{\rm{s}}(x)$. This inhomogeneity will lead to formation of a space charge $\rho_{\rm{e}}$ outside the EDL. However, in our discussion of numerical results we will show that this space charge is small. By treating the concentration profile as a collection of resistors in series\cite{coupled1} we can obtain the ultimate, laterally constant current. From this it follows that $\bar{\rho}_{\rm{s}}(x)$ in Eq.~(\hyref{cond}) should be replaced by the inverse average $L/\int_0^L(\bar{\rho}_{\rm{s}}(x)^{-1}\dd x$ which is close to the lateral average $\langle\bar{\rho}_{\rm{s}}\rangle$ as long as $|\log(\bar{\rho}_{\rm{s}}(x)/(2\rho_{\rm{b}})|\textless 1$. The error of this approximation diverges when $\bar{\rho}_{\rm{s}}(x)/2\rho_{\rm{b}}$ approaches zero. As in the Letter we restrict attention to concentration profiles that deviate less than about 30\% from the bulk concentration a regime for which this approximation is very reasonable.\\

Having already calculated $\mathbb{L}_{12}$ for the electro-osmotic flow $Q_{\rm{\psi}}$ we can invoke Onsager's reciprocal relation, which states that $\mathbb{L}_{21}=\mathbb{L}_{12}$\cite{linresponse1, linresponse2}, to find the fully advective pressure-driven current and obtaining the full current
\begin{equation}
    I(\Delta \psi,\Delta P)=\frac{\pi R_{\rm{t}} R_{\rm{b}}}{L}\bigg(\frac{e^2D\Delta\psi}{k_{\rm{B}}T }\langle\bar{\rho}_{\rm{s}}\rangle-\Delta P \frac{\epsilon \psi_0}{\eta}\bigg).
    \label{electric current sup}
    \tag{\hyref{Stokes}1}
\end{equation}
Finally, combining $\mathbb{L}_{11}$, $\mathbb{L}_{12}$ and $\mathbb{L}_{22}$ our ultimate expression for the Onsager-like matrix Eq.~(1) reads
\begin{equation}
      \frac{\pi R_{\rm{b}}R_{\rm{t}}}{L}
    \begin{pmatrix}
     \dfrac{R_{\rm{b}}^2R_{\rm{t}}^2}{8\eta\langle R ^2\rangle} &  \dfrac{-\epsilon\psi_0}{\eta} \\
     -\dfrac{\epsilon\psi_0}{\eta} & \dfrac{e^2D}{k_{\rm{B}}T}\langle \bar{\rho}_{\rm{s}}\rangle
    \end{pmatrix}
    \begin{pmatrix}
    \Delta P\\
    \Delta \psi
    \end{pmatrix}
    =\begin{pmatrix}
    Q\\
    I
    \end{pmatrix}.
    \label{Onsager}
    \tag{\hyref{Stokes}2}
\end{equation}\\
  \section{Ideal pore geometry}
As the deviation of Ohmic current is largest when the difference between the laterally-averaged concentration  $\langle\bar{\rho}_{\rm{s}}\rangle$ and bulk concentration $2\rho_{\rm{b}}$ is largest it is interesting to note that an analytic expression for $\langle\Delta\bar{\rho}_{\rm{s}}\rangle=\langle\bar{\rho}_{\rm{s}}\rangle-2\rho_{\rm{b}}$ is available in the limit $\text{Pe}=0$, which is the limit near which the concentration difference is largest,
\begin{equation}
    \langle\Delta\bar{\rho}_{\rm{s}}\rangle=\dfrac{e\Delta\psi}{k_{\rm{B}}T}\frac{\sigma}{R_{\rm{t}}}\dfrac{\dfrac{R_{\rm{t}}}{R_{\rm{b}}}\bigg(2(\dfrac{R_{\rm{t}}}{R_{\rm{b}}}-1\big)-\big(1+\dfrac{R_{\rm{t}}}{R_{\rm{b}}})\log\big(\dfrac{R_{\rm{t}}}{R_{\rm{b}}}\big)\bigg)}{(1-\dfrac{R_{\rm{t}}}{R_{\rm{b}}})^2}. \label{davrho}
    \tag{\hyref{Stokes}3}
\end{equation}
The prefactor $(e\Delta\psi/k_{\rm{B}}T)(\sigma/R_{\rm{t}})$ is the tip Duhkin number times the dimensionless potential drop and bulk concentration, which diverges for vanishing tip radius, indicating that for the maximum non Ohmic conductivity the tip size should be as small as possible. However as we assumed a thin-EDL limit from the very beginning this prediction only remains valid as long as $ R_{\rm{t}}\gg \lambda_{\rm{D}}$. We estimate that optimization of non-linear current by minimization of the tip radius holds up to $R_{\rm{t}}\approx 10\lambda_{\rm{D}}$. It is easily checked that for fixed tip radius $\langle\Delta\bar{\rho}_{\rm{s}}\rangle$ of Eq.~(\hyref{davrho}) has a maximum at $R_{\rm{t}}/R_{\rm{b}}\simeq 0.22$, a geometry which hence optimizes diodic behavior. Furthermore, while $\Delta\rho$ as defined by Eq.~(5) in the main text is a measure for the concentration profile the ratio  $\Delta\rho/\langle\bar{\rho}_{\rm{s}}\rangle$ is large, with the proportionality constant between the two at zero flow being given by 
\begin{equation}
  \frac{\Delta\rho}{\langle\Delta\bar{\rho}_{\rm{s}}\rangle}=\dfrac{2(R_{\rm{t}}/R_{\rm{b}})^{-2} (1-\dfrac{R_{\rm{t}}}{R_{\rm{b}}})^3}{2\dfrac{R_{\rm{t}}}{R_{\rm{b}}}-2-(1+\dfrac{R_{\rm{t}}}{R_{\rm{b}}})\log(\dfrac{R_{\rm{t}}}{R_{\rm{b}}})},\tag{S14}
  \label{propconst}
\end{equation}  which for straight pores with $R_{\rm{t}}\simeq R_{\rm{b}}$ equals 12 and in the limit $R_{\rm{t}}/R_{\rm{b}}\ll 1$ is well approximated by $\Delta\rho/\langle\Delta\bar{\rho}_{\rm{s}}\rangle\simeq 2R_{\rm{b}}^2/ R_{\rm{t}}^2 \log(R_{\rm{b}}/R_{\rm{t}})$ which diverges for large base radii. Hence for conical pores in general $\Delta\rho\gg 12\langle\bar{\rho}_{\rm{s}}\rangle$ and for our standard parameter set in the main text $\Delta\rho/\langle\Delta\bar{\rho}_{\rm{s}}\rangle\simeq 100.9$ while for the optimal tip-to-base ratio $\Delta\rho/\langle\Delta\bar{\rho}_{\rm{s}}\rangle\simeq 68.3$. In principle one could absorb the proportionality constant Eq.~(\hyref{propconst}) in the definition of $\Delta\rho$ to obtain a measure that accurately represents the laterally-averaged concentration profile $\langle\Delta\bar{\rho}_{\rm{s}}\rangle$.\\ 

\begin{figure}[t!]
  \includegraphics[width=0.9\linewidth]{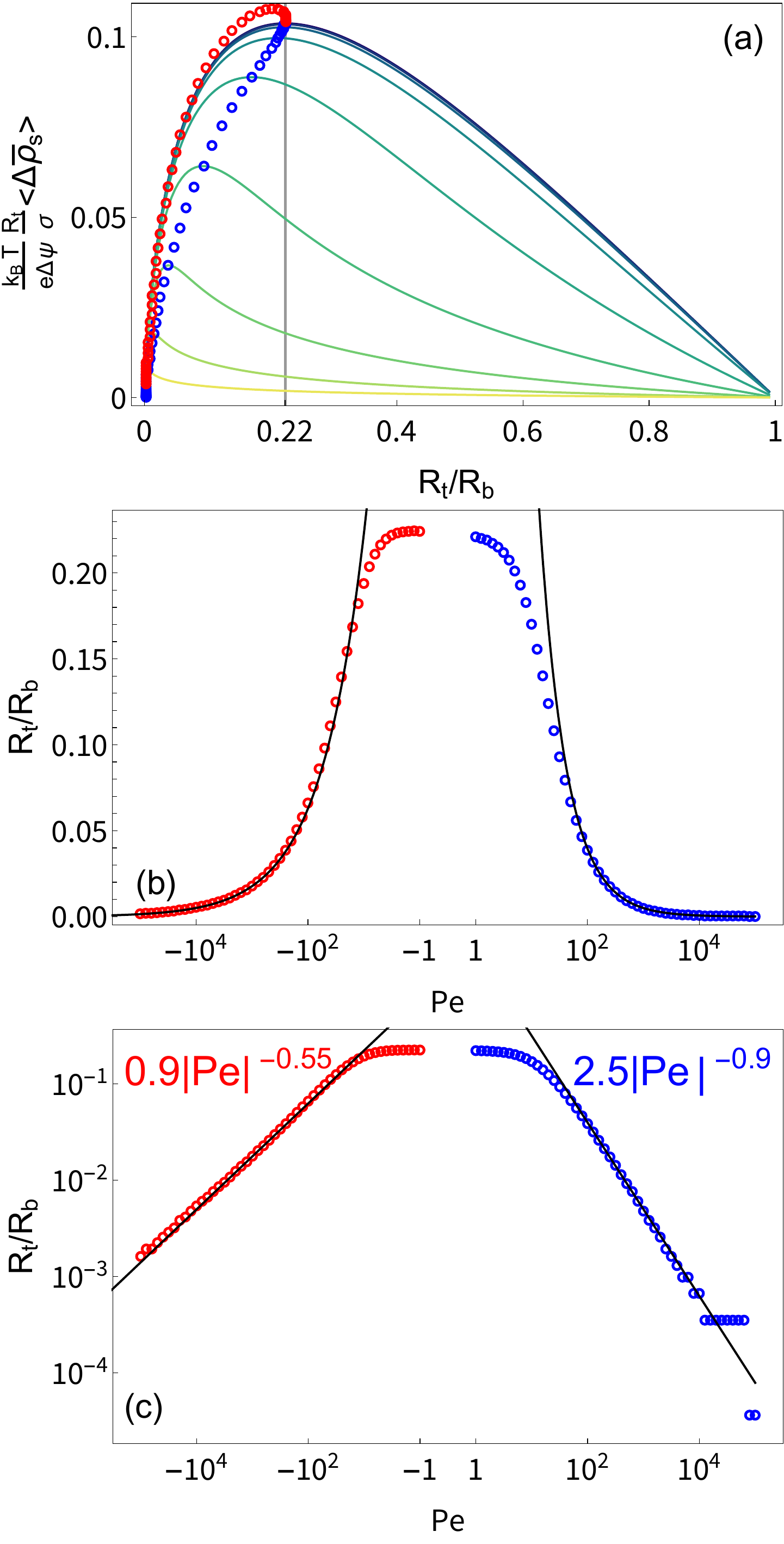}
  \caption{(a) Laterally averaged concentration $\langle\bar{\rho}_{\rm{s}}\rangle$ normalized by $(\sigma/R_{\rm{t}}) (e\Delta\psi/k_{\rm{B}}T)$ for varying tip-base ratio $R_{\rm{t}}/R_{\rm{b}}$ with the P\'eclet number between curves varying by $10^{1/2}$ with the yellow curve corresponding to Pe$=10^3$ and blue curve with Pe=$10^{-1}$ closely matching Eq.~(\hyref{davrho}). Blue points denote optimal tip-to-base ratios for Pe$\in[10^{-1},10^{5}]$ and red points denote optima for Pe$\in[-10^{-1},-10^{5}]$, with optimal ratios $R_{\rm{t}}/R_{\rm{b}}\ll1$ corresponding to Pe$\gg 1$.  (b) Optimal tip-to-base ratio for varying P\'eclet in log-linear representation, with red points at negative Pe and blue at positive Pe. Black lines represent power-laws whose scaling is found was found in (c) by fitting the data in log-log representation. For large positive Pe the optimal tip-to-base ratio scales as $R_{\rm{t}}/R_{\rm{b}}\simeq 2.5$|Pe|$^{-0.9 }$ while for large negative Pe it is well approximated by $R_{\rm{t}}/R_{\rm{b}}\simeq 0.9$|Pe|$^{-0.55}$.}
\label{fig:optimal}
\end{figure}

At finite Pe no analytic expression for $\langle\Delta\bar{\rho}_{\rm{s}}\rangle$ can be found, but its value can be straightforwardly calculated by numerical integration. In Fig.~\hyref{fig:optimal}(a) we plot $[(k_{\rm{B}}T R_{\rm{t}})/(e\Delta\psi\sigma)] \langle\Delta\bar{\rho}_{\rm{s}}\rangle$ against $R_{\rm{t}}/R_{\rm{b}}$ for Pe $\in[0,10^3]$. At Pe = 0 (blue line) the maximum laterally averaged concentration is found at a tip-to-base ratio $R_{\rm{t}}/R_{\rm{b}}\simeq 0.22$ (vertical line). As observed in Eq.~(\hyref{davrho}) for non-zero P\'eclet number the ideal tip-to-base ratio (symbols) is always smaller than $0.22$. Note that when no pressure drop $\Delta P^*$ is applied the P\'eclet number will scale with $\Delta\psi$ and the ideal pore geometry thus depends on the voltage operating range of a device. In Fig.~\hyref{fig:optimal}(b) we plot the  optimal tip-to-base ratio $R_{\rm{t}}/R_{\rm{b}}$ against P\'eclet number on a linear-logarithmic scale. It can be seen that for small |Pe|$\leq 10$ the ideal ratio 0.22 holds, but for large Pe it decays algebraically to zero. In Fig.~\hyref{fig:optimal}(c) we plot the same data as in (b) but now in a log-log representation to highlight the scaling in the Pe $\geq 100$ regime. We find that the relation between optimal geometry and Pe is well described by a power law $b|\text{Pe}|^{-\nu}$ in this regime, with $b\simeq 2.5$, $\nu\simeq0.9$ for positive and $b\simeq 0.9$, $\nu\simeq0.55$ for negative P\'eclet. 

\section{Discussion numerical results}

\begin{figure}[t!]
   \includegraphics[width=0.9\linewidth]{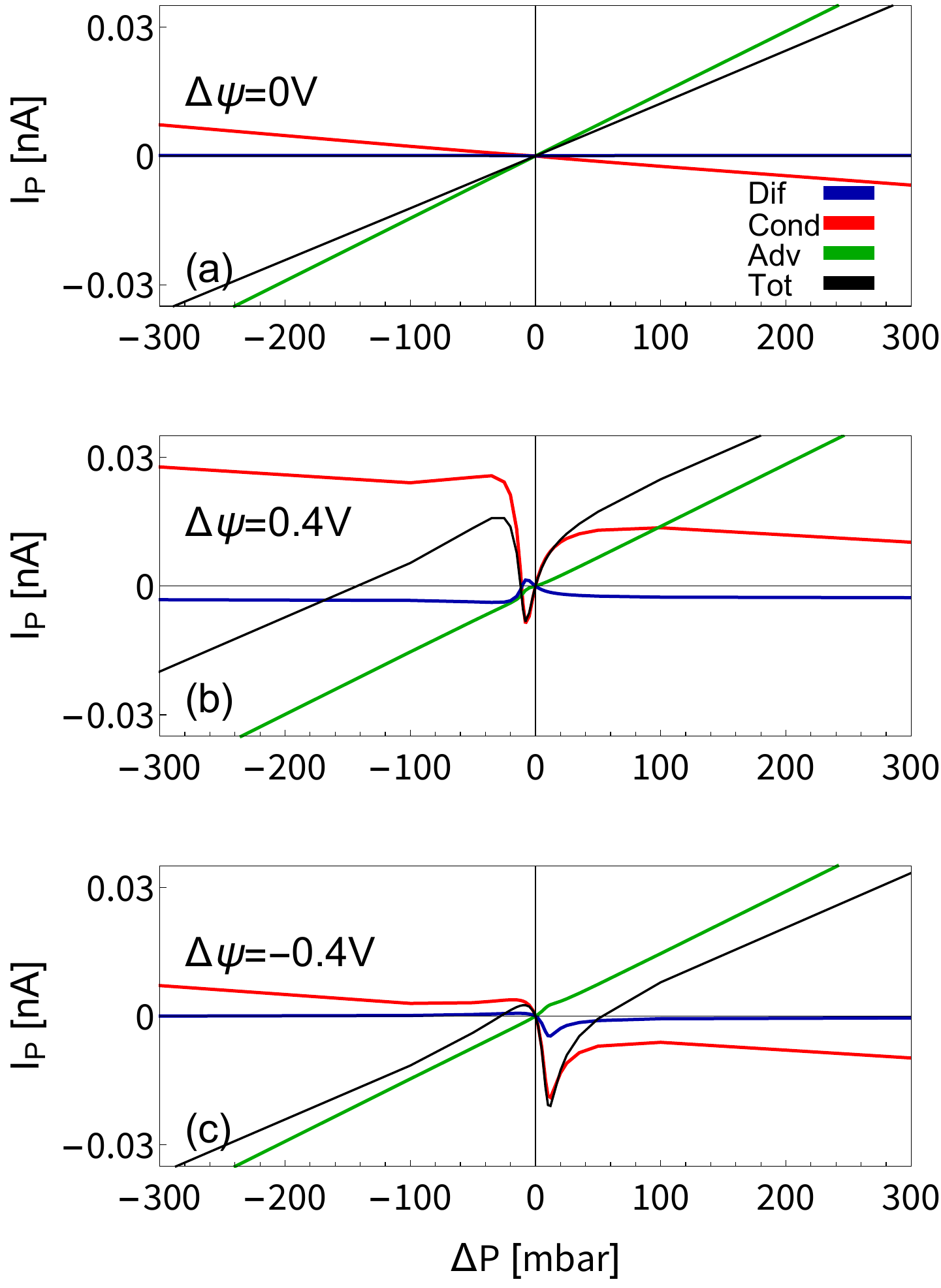}
  \caption{Total pressure excess diffusive, conductive, advection and net electric current (respectively Dif, Cond, Adv and $I_{\rm{P}}$) at $x=0.9L$ from numerical solutions of the full PNPS equations (\hyref{Stokes})-(\hyref{salt}) as function of the pressure drop $\Delta P$ offset by the current at $\Delta P=0$ for our standard set of parameters in the main text. For (a) where $\Delta\psi=0$ the current is linear in $\Delta P$ and dominated by the advective current. For (b) and (c), where respectively $\Delta\psi=+0.4$V and $\Delta\psi=-0.4$V, the current is non-linear for low pressure drops ($|\Delta P|$<50 mbar) and here conductive current dominates the non-linear pressure-current relation, with diffusive and advective components only marginally contributing. At large pressures the net current is again dominated by the advective, streaming current, current which follows the same linear trend found for $\Delta\psi=0$.}
\label{fig:currentcontribution}
\end{figure}
\begin{figure}[h!]
\includegraphics[width=0.45\textwidth]{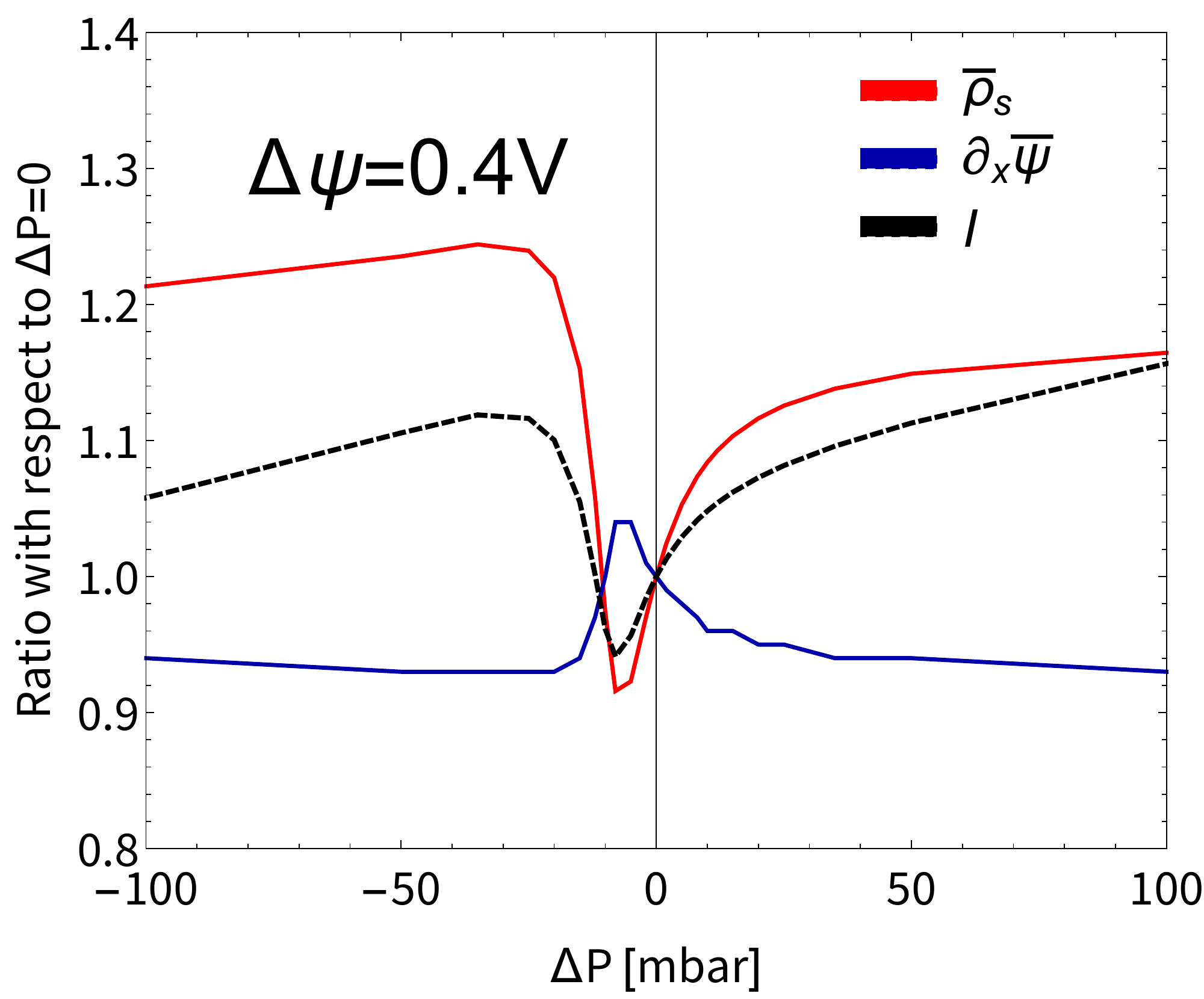}
\centering
\caption{Pressure drop dependence of the cross-sectional averaged electric field
$\partial_x\bar{\psi}$, salt concentration $\bar{\rho}_{\rm{s}}$, and total current $I$, all normalized with their respective values at $\Delta P=0$, at lateral position $x=0.9L$ and for a potential $\Delta\psi=0.4$V, as obtained  from numerical solutions of the full PNPS equations (\hyref{Stokes})-(\hyref{salt}) for our standard parameter set (see main text). The relative deviations from unity are much larger for the salt concentration and the current than for the electric field; in fact the current correlates well with the salt concentration and even anti-correlates with the electric field.}
\label{fig:evsconc}
\end{figure}

Here we will discuss the numerical results of the full PNPS equations (\hyref{Stokes})-(\hyref{salt}) in depth and show that the effect of the space charge outside the EDL on the current $I$ can be neglected.  In Fig.~\hyref{fig:currentcontribution} we plot our numerical solutions of the pressure excess electric current $I_{\rm{P}}=I(\Delta P,\Delta\psi)-I(0,\Delta\psi)$ as a function of the pressure drop $\Delta P$ for our standard parameter set and (a) $\Delta\psi=0$, (b) $\Delta\psi=+0.4$V, and (c) $\Delta\psi=-0.4$V. When no potential drop is applied we find that the electric current is linear in the pressure drop and dominated by advection, as can be seen in Fig.~\hyref{fig:currentcontribution}(a) where we plot $I_{\rm{P}}(\Delta P,\Delta\psi)$, split into its diffusive, conductive and advective components. When the pressure drop is applied in conjunction with a potential drop we find that it is strongly non-linear, as can be seen Fig.~\hyref{fig:currentcontribution}(b) and (c).  A minimum in the total current is found at $\Delta P=\mp 10$ mbar for $\Delta \psi=\pm 0.4$V in Fig.~\hyref{fig:currentcontribution}(b) and (c), and near the minimum the conductive current dominates over the advective and diffusive currents. For pressure drops larger than $|\Delta P|$>100 mbar the advective current $I_{\rm{P}}$ dominates and follows the linear relation observed for $\Delta\psi=0$ shown in Fig.~\hyref{fig:currentcontribution}(a). The non-linearities in (b) and (c) are mainly due to the conductive component $-(De/k_BT)\rho_{\rm{s}}(x,r)\partial_x\psi(x,r)$ of the electric current Eq.~(\hyref{charge}) which depends not only on the salt concentration $\rho_{\rm{s}}(x,r)$ but also the electric field $-\nabla\psi(x,r)$.  Inspecting Fig.~\hyref{fig:evsconc}, where we plot the $\Delta P$-dependence of the current, the salt concentration, and the electric field at $x=0.9L$ all normalized by their values at vanishing pressure drop, we indeed find that both the salt concentration $\bar{\rho}_{\rm{s}}$ as well as the electric field $-\partial_x\bar{\psi}$ vary with $\Delta P$. However, we find that the variation of the electric field is $\sim 5$ times smaller than the change of concentration with pressure, and actually opposes the non-linearity of the conduction current for $\Delta\psi=-0.4$V.

As the change in electric field actually counter-acts the non-linearity observed in the electric current we conclude that the $\Delta P$ dependency of $\partial_x\bar{\psi}$ cannot be the dominant driving force behind the non-linear current $I(\Delta P)$. This suggests that the non-linear current $I$ can be essentially understood by considering the dependency of the salt density $\rho_{\rm{s}}$ on $\Delta\psi$ and $\Delta P$, with the space charge density $\rho_{\rm{e}}$ outside of the EDL only contributing minutely to the non-linearity both through diffusive $I_{\rm{dif}}$ and advective $I_{\rm{adv}}$ currents as well as pressure dependent electric field $-\partial_x\bar{\psi}$. Hence Fig.~\hyref{fig:evsconc} shows that dependency of space charge $\rho_{\rm{e}}$ on $\Delta P$ only leads to negligible variation in the magnitude of diffusive and advective current $I_{\rm{dif}}$ and $I_{\rm{adv}}$ as seen in Fig.~\hyref{fig:currentcontribution}(b) and (c). The conclusion that space charge is negligible is based on an empirical observation of numerical results, so we cannot rule out that there are regimes in which space charge does dominate the non-linear current. However, as we have chosen a set of parameters to reproduce the experimental set-up of Ref.~\cite{jubin} we can conclude that for this specific set of experiments space charge is negligible and not of key importance to the non-linear current $I(\Delta P)$. We conclude from these numerical results that (i) the conductive component of the electric current is responsible for the extremely mechanosensitive current observed in experiments,  (ii) changes in the electric field and space-charge density with pressure are small and can be neglected, and (iii) the pressure-sensitivity of the electric-field and space charge actually (weakly) oppose the non-linear conduction current and thus cannot be responsible for the observed non-linear current.\\  

	\bibliographystyle{apsrev4-2}